\documentclass[11pt,a4paper]{article}
\pdfoutput=1
\usepackage{jcappub}
\usepackage{natbib}
\usepackage{epsfig,multicol,bbm}
\usepackage{graphicx}
\usepackage{dcolumn}
\usepackage{amsmath}
\usepackage{enumerate}
\usepackage{color}
\usepackage[margin=2.2cm, top=1.9cm, bottom=2.5cm]{geometry}
\def\aa{{\sl Astron.\ \&\ Astrophys.\ }}

\def\apj{{\sl Astrophys.\ J.\ }}
\def\apjl{{\sl Astrophys.\ J.\ Lett.\ }}
\def\apjs{{\sl Astrophys.\ J.\ Supp.\ }}

\def\jcap{{\sl J.\ Cosm.\ Astroparticle\ Phys.\ }}
\def\jltp{{\sl J.\ Low. \ Temp.\ Phys.\ }}

\def\mnras{{\sl MNRAS\ }}
\def\mpla{{\sl Mod.Phys.Lett.A\ }}

\def\physrep{{\sl Phys.\ Rept.\ }}
\def\plb{{\sl Phys.\ Lett.\ B\ }}

\def\prd{{\sl Phys.\ Rev.\ D\ }}
\def\prl{{\sl Phys.\ Rev.\ Lett.\ }}

\def\pdu{{\sl Phys.\ Dark\ Univ.}}

\newcommand{\lsim}{\,\lower2truept\hbox{${<\atop\hbox{\raise4truept\hbox{$\sim$}}}$}\,}
\newcommand{\gsim}{\,\lower2truept\hbox{${>\atop\hbox{\raise4truept\hbox{$\sim$}}}$}\,}

\newcommand{\etal}{{et\ al.}}

\title{Maps of CMB lensing deflection from N-body simulations in Coupled Dark Energy Cosmologies}

\vspace{0.5cm}
\author[a,c,d]{Carmelita Carbone}
\author[b,c,e]{Marco Baldi}
\author[f]{Valeria Pettorino}
\author[g]{Carlo Baccigalupi}

\vspace{0.5cm}
\affiliation[a]{INAF -- Osservatorio Astronomico di Brera,
Via Bianchi 46, I-23807 Merate (LC), ITALY}
\affiliation[b]{Dipartimento di Fisica e Astronomia, Universit\`{a} di Bologna, Viale B.~Pichat 6/2, I-40127 Bologna, Italy}
\affiliation[c]{INAF -- Osservatorio Astronomico di Bologna, via
  Ranzani 1, I-40127 Bologna, Italy}
\affiliation[d]{INFN, Sezione di Bologna, Viale Berti Pichat 6/2,
  I-40127 Bologna, Italy}
\affiliation[e]{Excellence Cluster Universe
Boltzmannstrasse 2, 85748 - Garching bei M\"unchen, Germany}
\affiliation[f]{D\'epartement de Physique Th\'eorique and Center for Astroparticle Physics, Universit\'e de Gen\`eve, 24 quai Ernest Ansermet, CH--1211 Gen\`eve 4, Switzerland}
\affiliation[g]{SISSA, Via Bonomea 265, Trieste, I-34136, Italy}

\emailAdd{carmelita.carbone@brera.inaf.it}
\emailAdd{marco.baldi5@unibo.it}
\emailAdd{valeria.pettorino@unige.ch}
\emailAdd{bacci@sissa.it}

\vspace{0.5cm}
\abstract{We produce lensing potential and deflection-angle maps in order to simulate the weak gravitational lensing of the Cosmic Microwave Background (CMB)
via ray-tracing through the COupled Dark Energy Cosmological
Simulations ({\small CoDECS}), the largest suite of N-body simulations
to date for interacting Dark Energy cosmologies.  
The constructed maps faithfully reflect the N-body cosmic structures on a range of scales going from the arcminute to the degree scale, limited only by the resolution and extension 
of the simulations. We investigate the variation of the lensing
pattern due to the underlying Dark Energy (DE) dynamics, characterised
by different  
background and perturbation behaviours as a consequence of the
interaction between the DE field and Cold Dark Matter (CDM). In
particular, we study in detail the results from three cosmological
models differing in the background and perturbations evolution at the 
epoch in which the lensing cross section is most effective,
corresponding to a redshift of $\sim 1$, with the purpose to isolate their 
imprints in the lensing observables, regardless of the compatibility
of these models with present constraints. The scenarios investigated
here include a reference $\Lambda$CDM cosmology, a
standard coupled DE (cDE) scenario, 
and a ``bouncing" cDE scenario.
For the standard cDE scenario, we find that typical differences in the
lensing potential result from two effects: the enhanced growth of linear CDM
density fluctuations with respect to the $\Lambda$CDM case, and the
modified nonlinear dynamics of collapsed structures induced by the DE-CDM interaction.
As a
consequence, CMB lensing highlights the DE impact in the cosmological
expansion, even in the degenerate case where the amplitude of the linear matter density perturbations, parametrised through
$\sigma_{8}$, is the same in both the standard cDE and $\Lambda$CDM
cosmologies. For the ``bouncing" scenario, we find that the two
opposite behaviours of the lens density 
contrast and of the matter abundance lead to a counter-intuitive
effect, making the power of the lensing signal in this model lower
by $10\%$ than in the $\Lambda$CDM scenario. 
Moreover, we compare the behaviour of CDM and baryons in {\small
CoDECS} separately, 
in order to isolate effects coming from the coupling with the DE
component. We find that, in the bouncing scenario, baryons show an
opposite trend with respect to CDM, due to the coupling of the
latter with the DE component.
These results confirm the relevance of CMB lensing as a probe for DE
at the early stages of cosmic acceleration, and demonstrate the
reliability of N-body based large scale CMB lensing 
simulations in the context of DE studies.}

\keywords{lensing, CMB, dark-energy, dark-matter, cosmology}

\begin{document}
\maketitle
\section{Introduction}
\label{sec:introduction}
Anisotropies in the Cosmic Microwave Background (CMB) represent a pillar of our present understanding of cosmology. They are sourced by 
all kinds of cosmological perturbations, namely scalar (such as density), vector (vorticity) and tensor (Gravitational Waves, GWs). At recombination,
Thomson scattering stores imprints of such perturbations in the intensity and linear polarisation of photons and enables us to observe them as 
primary anisotropies in the CMB sky, to be distinguished from the
secondary ones collected along the line of sight. The scattering produces total 
intensity ($T$) anisotropies and linear polarisation ones, described by the Stokes parameters $Q$ and $U$, see e.g. \citet{hu_etal_1998} for reviews. 
The latter are further combined into gradient modes (or electric-type,
$E$) and curl ones (or magnetic-type, $B$). At the linear level, while
the $E$ modes are  
sourced by all kinds of perturbations, $B$ modes are excited only by
vorticity and
GWs \citep{kamionkowski_etal_1997,zaldarriaga_seljak_1997}. \\  
The $T$ power on large scales, greatly dominated by the metric effect
known as Sachs-Wolfe (SW), occupies angular scales up to the angle
subtended by the  
horizon at recombination, corresponding to a multipole
$l\simeq 200$ in spherical harmonics; degree and sub-degree angular
scales are dominated by acoustic oscillations  
from radiative pressure opposing gravity. Polarisation is caused by
local anisotropy in $T$ on the last scattering region, and has
significant power on the degree and  
sub-degree scales, featuring acoustic oscillations, largely correlated
with those in $T$.  

The Planck satellite \citep{planck2013-p01} has observed CMB anisotropies for $T$ over the whole sky down to a resolution of 5 arcminutes, with a sensitivity 
of a few $\mu$K per resolution element, observing large scales and
seven acoustic oscillations. Measurements of the $TE$ correlation have
also been reported \citep{planck2013-p01} and the analysis of data is
still ongoing, in particular the polarisation  
measurements  will be made public in a forthcoming release. 
Previously, the Wilkinson Microwave Anisotropy Probe (WMAP) reached an angular resolution of about $20$ arcminutes, 
also producing measurements of the $TE$ correlation \citep[see e.g.][
and references therein]{bennett_etal_2012}. Operating experiments from
the ground and monitoring percent 
sky fractions have already reported the existence of acoustic oscillations for $T$ down to the arcminute scale, namely the South Pole Telescope 
\citep[SPT, see][]{story_etal_2012,keisler_etal_2011} and the Atacama Cosmology Telescope \citep[APT, see][]{hlozek_etal_2012,das_etal_2011}. 
Sub-orbital probes such as the $E$ and $B$ experiment\footnote{groups.physics.umn.edu/cosmology/ebex} \citep[EBEX, see][]{reichborn_kjennerud_etal_2011} and POLARBEAR 
\footnote{bolo.berkeley.edu/polarbear} \citep{arnold_etal_2010}, Spider \citep{mactavish_etal_2008}, SPTpol \citep{bleem_etal_2012a}, ACTpol \citep{niemack_etal_2010}, 
the Large Scale Polarisation Experiment \citep[LSPE,
see][]{bersanelli_etal_2012}, QUbic \citep{piat_etal_2012} are
designed to probe CMB polarisation looking in particular  
for the $B$ modes from primordial GWs on percent sky fractions with angular resolution similar to Planck; moreover, proposals for a future post-Planck CMB satellite, aiming 
at an all sky study of CMB polarisation, are in preparation: see e.g. the Cosmic ORigin Explorer\footnote{www.core-mission.org} (CORE), CMBpol\footnote{cmbpol.uchicago.edu}, 
the Primordial Inflation Explorer \citep{kogut_etal_2011}, and the Light (Lite) satellite for the studies of $B$-mode polarisation and Inflation from cosmic background Radiation Detection 
(LiteBIRD\footnote{http://cmbpol.kek.jp/litebird}). 

The data available at present are well represented by a six parameter cosmological model known as $\Lambda$CDM \citep[see][ and references therein]{planck2013-p11}: the expansion proceeds 
with a rate of $\sim 68$ km/s/Mpc, and is accelerating under the
effect of a Dark Energy (DE) component constituting $\sim 68\%$ of
the entire cosmic energy density budget, the rest being described 
as Cold Dark Matter (CDM) responsible for the dark halos around galaxies and galaxy clusters (about $27\%$) as well as leptons and baryons (about $5\%$); the primordial density power spectrum 
departs from scale invariance, $\propto k^{n_{s}}$, where $n_{s} \sim
0.96$. No detection of cosmological GWs exists to date.  

Present and future CMB observations are constraining the
cosmological dark components, through the interaction between the CMB and the forming cosmological Large Scale Structures (LSS) along 
the line of sight. This latter aspect represents the context of the present work.

Secondary anisotropies are caused by the interaction of CMB photons
with LSS along their paths from the last scattering surface to
the observer. CMB photons re-scatter onto electrons freed by cosmological 
reionisation at the beginning of structure formation. This boosts polarisation anisotropies on the angular scales subtended by the horizon at that epoch, corresponding to several degrees in the sky, 
as it has been detected by the WMAP satellite \citep[see e.g.][ and references therein]{bennett_etal_2012}. Moreover, the Integrated SW (ISW) effect is caused by CMB photons crossing regions 
characterised by evolving gravitational potentials, and has been now observed by Planck \citep{planck2013-p14}; this effect is active already for linearly evolving LSS while it is known as Rees-Sciama (RS) 
when the underlying structures are non-linear. The efficiency of the ISW reconstruction from data is boosted when CMB and LSS are Cross Correlated (XC); several authors have used 
techniques aiming at estimating such a correlation in order to extract 
cosmological constraints on DE as well as on the statistics of primordial perturbations from the ISW
\citep{ho_etal_2008,giannantonio_etal_2008,xia_etal_2011,1010.2192}. 

The subject of this work concerns a third aspect of secondary
anisotropies, namely the capability of LSS to act as gravitational lenses for CMB photons. 
The combination of the statistical spatial distribution of power in
the CMB anisotropies and LSS, associated with the geometry of the expansion, makes the CMB lensing process 
effective on sub-degree and arcminute angular scales, causing a smearing of acoustic peaks and the transfer of power in the damping tail of CMB primary anisotropies, as well as a transfer of the $E$ mode 
power into the $B$ modes, resulting in a characteristic peak in the
latter at multipoles of about $l \sim 1000$ \citep{zaldarriaga_seljak_1998}. Planck has measured the CMB lensing with about 
$26\sigma$ confidence level, covering the whole spectrum of scales
where it is effective \citep{planck2013-p12}, along its correlation with the structures observed as the Cosmic Infrared Background 
\citep[CIB, see][ and references therein]{planck2013-p13}. Moreover, previously and independently, the ACT \citep{das_etal_2011} 
and SPT \citep{keisler_etal_2011} experiments detected the lensing
signal in the damping tail of CMB total intensity anisotropies. 

The lensing cross section 
peaks half way between the observer and the source. If the source is
effectively at infinity as in the case of CMB, the epoch in which the
lensing distortion is most recorded into CMB anisotropies coincides
with the onset of cosmic acceleration, at $z\simeq 1\pm 0.5$:
therefore, since the lensing efficiency is related to the behaviour of
the  
expansion and perturbations at the corresponding epoch, the CMB
lensing signal carries important information about the onset of
late-time cosmic acceleration, poorly constrained at present, promoting 
CMB {\it alone} to be a probe of the DE behaviour at redshift of about 1, {\it regardless of how close it is to a Cosmological Constant at present} \citep{acquaviva_baccigalupi_2006}. 
On the contrary, primary CMB anisotropies suffer the projection
degeneracy making them unable 
to distinguish a global curvature from a Cosmological Constant. Predictions say that CMB lensing should be able to measure the DE abundance at the
onset of acceleration with $\sim 10\%$ precision \citep{acquaviva_baccigalupi_2006,hu_etal_2006}. ACT and and SPT data have already been used 
\citep{sherwin_etal_2011,van_engelen_etal_2012} in combination with other cosmological data sets, to break the projection degeneracy and give evidence of 
DE from the CMB alone. Recent theoretical works are investigating the
capabilities of CMB lensing to measure not only the abundance of DE at
the onset of acceleration, but its  
coupling with CDM and of early DE \citep{1301.5279}.
The CMB lensing as a DE probe will benefit from the XC with the observations of the actual lenses in LSS surveys. The XC with CMB lensing has been already detected on the basis of the available CMB and LSS data \citep{bleem_etal_2012b,sherwin_etal_2012}. The latter will culminate in about a decade with 
the observations of the Euclid satellite\footnote{www.euclid-ec.org}, which will perform arcsecond imaging of billions of galaxies over half of the sky between redshifts 0 and 2, 
with photometric redshift accuracy corresponding to the percent level,
reaching $0.1\%$ from spectroscopic measurements for a sub-sample of
millions of them for $0.7<z<2.1$ \citep{1110.3193,1206.1225}. The mapping of the galaxy clustering on such a volume, as well as the weak lensing shear associated to the induced ellipticity in galaxy images, represent Euclid primary targets in order to constrain the DE 
behaviour in the corresponding redshift interval. 

In order to be prepared to fully exploit the potential of the
observations outlined above for the investigation of the dark
cosmological components, an intense preparatory work aiming at a 
detailed simulation of the interplay between structure formation and CMB lensing is necessary. The starting point is the evaluation of the lensing pattern through 
the N-body simulations of structure formation. This has the potentiality to reproduce the full statistics of the lensing signal, relaxing the assumption of Gaussianity beyond 
the two point correlation function
\citep{carbone_etal_2008,fosalba_etal_2008,carbone_etal_2009,teyssier_etal_2009}; in these works, the signal for the power
spectrum, predicted semi-analytically, was correctly recovered, while all sky maps of deflection angle and projected lensing potential were constructed, as well as lensed CMB 
$T$, $E$ and $B$ anisotropies, down to an angular resolution of a few arcminutes. Similar studies, consisting in the simulation of the ISW through N-body LSS, have also been performed 
\citep{cai_etal_2010}. 

In this work we make a further progress along this direction. For the first time we conduct ray tracing studies through N-body simulations of non-$\Lambda$CDM cosmologies \citep[for a recent review on N-body simulations in DE cosmologies, see][]{Baldi_2012c}. 
The main focus of our analysis is the study and characterisation of the response of lensing to the underlying DE behaviour, focusing on the epoch at which the lensing 
is most effective. As representatives of various DE cosmologies, we
adopt the ones implemented in the {\small
CoDECS}\footnote{www.marcobaldi.it/CoDECS} publicly available
suite of N-body simulations \citep[][]{CoDECS}. The latter include various realisations
of dynamical DE models characterised by a direct interaction between  
DE and CDM particles. 

This work is organised as follows. In \S~\ref{sec:codecs} we describe
the set of investigated cosmologies and their implementation through
the CoDECS N-body simulations; in \S~\ref{sec:cmb_lensing} we review
the basics of CMB lensing and the approach we use to trace CMB photons 
through the CoDECS structures. In \S~\ref{sec:results} we show our
results. Finally, in \S~\ref{sec:conclusions} we summarise our work
and draw the concluding remarks. 
\section{DE cosmologies in CoDECS}
\label{sec:codecs}
As a first extension of CMB lensing studies to cosmological scenarios beyond $\Lambda $CDM, we focus on models where the role of DE is played by a classical 
scalar field $\phi$ dynamically evolving in a self-interaction
potential $V(\phi )$, and characterised by a direct interaction with
CDM particles. This class of cosmologies, 
generally referred to as ``coupled DE" models (cDE hereafter) have
been widely investigated in the literature as a possible way to 
alleviate the fine-tuning problems of the Cosmological
Constant \citep[see e.g.][and references therein]{Wetterich_1995, Amendola_2000, mangano_etal_2003, Amendola_2004,Farrar_Peebles_2004, Pettorino_Baccigalupi_2008}. 
At the background level, cDE models are described by an interaction
term between DE and CDM continuity equations, while baryons are kept uncoupled:
\begin{eqnarray}
\label{klein_gordon} 
\ddot{\phi } + 3H\dot{\phi } +\frac{dV}{d\phi } &=&
\sqrt{\frac{2}{3}}\beta _{c}(\phi ) \frac{\rho _{c}}{M_{{\rm Pl}}} \,, \\
\label{continuity_cdm}
\dot{\rho }_{c} + 3H\rho _{c} &=& -\sqrt{\frac{2}{3}}\beta _{c}(\phi
)\frac{\rho _{c}\dot{\phi }}{M_{{\rm Pl}}} \,, \\
\label{continuity_baryons}
\dot{\rho }_{b} + 3H\rho _{b} &=& 0 \,.
\end{eqnarray}
Here the coupling function $\beta _{c}(\phi )$ sets the strength of the interaction and the sign of the quantity $\dot{\phi }\beta_{c}(\phi )$ 
defines the direction of the energy flow between the DE and CDM fluids. 
Although in the most general case the coupling function can directly depend on the scalar field itself, in the present work we will focus on a subset of the {\small CoDECS} cosmological models characterised by a constant coupling $\beta _{c}(\phi ) = \beta _{c}$, and we will then omit the scalar field dependence of the coupling function in the remainder of the paper.
In Eqs.~(\ref{klein_gordon})-(\ref{continuity_baryons}) the
overdot represents the derivative with respect to cosmic time $t$ and $M_{\rm Pl}\equiv 1/\sqrt{8\pi G}$ is the reduced Planck mass,
with $G$ the Newton's constant. According to Eqs.~(\ref{klein_gordon})-(\ref{continuity_cdm}), different choices of the scalar field self-interaction potential $V(\phi )$ and coupling
 $\beta _{c}$ determine a different evolution of the DE density $\rho _{\phi }\equiv \dot{\phi }^{2}/2 + V(\phi )$ and
equation of state $w_{\phi }\equiv \left[ \dot{\phi }^{2}/2 - V(\phi   )\right] /\rho _{\phi }$, which allows to put broad constraints on
the model's parameters based on background observables only \citep[see e.g.][and references therein]{Amendola_2000, Bean_etal_2008, lavacca_etal_2009, Pettorino_etal_2012}. In
particular, cDE models are in general characterised by a scaling
regime in the matter dominated era during which DE provides a
fraction of the total cosmic energy  
density, directly proportional to the square of the coupling, $\Omega _{\phi }\approx 4\beta _{c}^{2}/3$. This regime is known as the $\phi $-Matter Dominated Epoch 
(or $\phi $-MDE, see \citet{Amendola_2000}). The DE-CDM interaction also determines a time evolution of the CDM particle mass given by the integral of Eq.~(\ref{continuity_cdm}): 
\begin{equation}
\label{mass_variation}
\frac{d\ln \left[ M_{c}/M_{\rm Pl}\right] }{dt} = - \sqrt{\frac{2}{3}}\beta _{c}\dot{\phi }\,,
\end{equation}
where $M_{c}$ is the mass of a CDM particle such that $\rho _{c} =
M_{c}n_{c}$, with the number density $n_{c}$ being constant in
comoving coordinates. 
Such a variation of the CDM particle mass induces in turn a shift of
the matter-radiation equivalence redshift $z_{\rm eq}$. All these
effects allow to constrain the coupling strength, in particular, for
the simplest class of models characterised by the constant couplings
$\beta _{c}$ considered in the present work. 

Besides providing a non-trivial evolution of the DE density and equation of state parameter, cDE models are also characterised by a long-range fifth-force, 
mediated by the DE scalar field and acting between CDM particles. In this respect, the relation between cDE cosmologies and modified gravity models, as e.g. scalar-tensor
theories of gravity, has been extensively discussed by e.g. \citet{Pettorino_Baccigalupi_2008}. The growth of linear density perturbations on sub-horizon scales can be described in 
the Fourier space and in the Newtonian gauge by the evolution equations for the CDM and baryon density contrast $\delta _{c,b}$: 
\begin{eqnarray}
\label{gf_c}
\ddot{\delta }_{c} &=& -2H\left[ 1 - \beta _{c}\frac{\dot{\phi }}{H\sqrt{6}}\right] \dot{\delta }_{c} + 4\pi G \left[ \rho _{b}\delta _{b} + \rho _{c}\delta _{c}\Gamma _{c}\right] \,, \\
\label{gf_b} 
\ddot{\delta }_{b} &=& - 2H \dot{\delta }_{b} + 4\pi G \left[ \rho
  _{b}\delta _{b} + \rho _{c}\delta _{c}\right]\,, 
\end{eqnarray} 
where the factor $\Gamma _{c}\equiv \left( 1 + 4\beta _{c}^{2}/3\right)$ includes the effect of the fifth-force mediated by the DE scalar field for CDM density perturbations. 
The second term in the first square bracket at the right-hand side of
Eq.~(\ref{gf_c}), instead, represents the velocity dependent term
arising for CDM perturbations as a consequence  
of momentum conservation, first described in detail in \citet{baldi_etal_2010}. Both these two additional terms contribute significantly to modify the evolution of density 
perturbations as compared to the $\Lambda $CDM cosmology and even to
an uncoupled scalar field DE model. In particular, while the
fifth-force is always attractive for the case -- considered in the present work -- of a single 
CDM particle species (while the same is in general not always true for multiple CDM families, see e.g. 
\citet{Brookfield_VanDeBruck_Hall_2008,Baldi_2012a,Baldi_2012b}), the friction term can act both as a proper friction or as a dragging effect, depending on the sign of the quantity 
$\dot{\phi }\beta _{c}$. The interplay between these two terms can imprint very peculiar features on the growth rate of matter density perturbations, allowing to 
put tighter constraints on the coupling and potential functions and to possibly reconstruct their functional form. At the non-linear level, the acceleration equation of coupled (CDM) 
particles is characterised by the same two additional terms appearing in Eq.~(\ref{gf_c}): 
\begin{equation} 
\dot{\vec{v}}_{c} = \beta _{c}\frac{\dot{\phi
}}{\sqrt{6}}\vec{v}_{c} - \vec{\nabla }\left[
  \sum_{c}\frac{GM_{c}(\phi )\Gamma _{c}}{r_{c}} +
  \sum_{b}\frac{GM_{b}}{r_{b}}\right] \,,
\end{equation} 
with the only further complication that the particle's velocity and acceleration are no longer necessarily aligned, thereby significantly increasing the complexity of the possible interplay
between the friction and fifth-force effects for the non-linear structure formation and evolution \citep[see e.g.][]{Baldi_2011b}. 

The {\small CoDECS} simulations \citep{Baldi_2012b}, which represent the largest suite of cosmological N-body and hydrodynamical simulations of cDE models to date, have been obtained 
by making use of a modification by \citet{baldi_etal_2010} of the widely used parallel N-body code GADGET \citep{springel_etal_2005} that allows to self-consistently simulate the evolution 
of structure formation processes in the context of generalised cDE cosmologies by including all the relevant features just discussed, from their specific background evolution to the mass 
variation of CDM particles, to the effects of the fifth-force and of the extra-friction acting on individual particles. Such modified code has been widely used in the past to investigate 
cDE cosmologies with both constant \citep{baldi_etal_2010,baldi_viel_2010,Baldi_2011b} and variable \citep{Baldi_2011a,baldi_pettorino_2011,Marulli_Baldi_Moscardini_2012,Giocoli_etal_2012} couplings, as well as for different choices 
of the scalar field self-interaction potential. All the {\small CoDECS} simulations have been recently made publicly available through a dedicated web
database\footnote{www.marcobaldi.it/CoDECS}. At present, they include two distinct sets of runs, the {\small L-CoDECS} and the {\small H-CoDECS}. The {\small L-CoDECS} simulations consist 
of cosmological boxes of $1$ comoving Gpc$/h$ aside, filled with $1024^{3}$ CDM and $1024^{3}$ baryon particles. Both types of particles are treated with collisionless dynamics only, which means 
that baryonic particles are not considered as gas particles but just as a different family of collisionless particles distinguished from CDM. This is done in order to account for the effect of 
the uncoupled baryon fraction in cDE models which would not be correctly represented by CDM-only simulations. The mass resolution at $z=0$ for this set of simulations is $M_{c}=5.84\times 10^{10}$ M$_{\odot }/h$ for CDM and $M_{b}=1.17\times 10^{10}$ M$_{\odot }/h$ for baryons, while the gravitational softening is set to $\epsilon_{s}=20$ comoving kpc/$h$, corresponding to $0.04$ times the
mean linear interparticle separation. The {\small H-CoDECS} simulations are instead adiabatic hydrodynamical simulations on much smaller scales, which we do not consider in the present work. 

The {\small CoDECS} suite presently includes six different cosmological scenarios: a fiducial $\Lambda $CDM cosmology taken as reference, three cDE models with a constant positive coupling 
$\beta_{c}>0$ and an exponential self-interaction potential:
\begin{equation}
\label{exponential}
V(\phi ) = Ae^{-\alpha \phi } \,,
\end{equation}
one cDE model with the same potential and an exponential coupling function:
\begin{equation}
\beta _{c}(\phi ) \equiv \beta _{0}e^{\beta _{1}\phi }\,,
\end{equation}
and one final cDE scenario with a constant negative coupling $\beta_{c}<0$ and a SUGRA \citep[][]{Brax_Martin_1999} self-interaction potential:
\begin{equation}
\label{SUGRA}
V(\phi ) = A\phi ^{-\alpha }e^{\phi ^{2}/2} \,.
\end{equation}
The exponential potential of Eq.~(\ref{exponential}) is known to provide, in the absence of couplings, stable attractor solutions characterised by a constant ratio between the DE density and that of
the dominant cosmological component \citep{liddle_scherrer_1999}. When a coupling is active, however, the transient meta-stable scaling regime already discussed above and known as $\phi $-MDE
occurs \citep{Amendola_2000}. The SUGRA potential form derives instead from supersymmetric theories of gravity \citep{brax_etal_2001}, determining a flattening of the potential at low redshifts in 
the case of no coupling; \citet{Baldi_2012b} has extensively investigated the same SUGRA model in the case of non-zero coupling \citep[see also][]{Tarrant_etal_2012}, showing that it determines a 
``bounce" of the DE equation of state $w_{\phi}$ on the cosmological constant barrier $w_{\phi }=-1$ at relatively recent times for initial conditions compatible with a static scalar field in the early
Universe, $\dot{\phi }(z\rightarrow \infty )=0$. Such cDE model,
characterised by a constant negative coupling and by a SUGRA
self-interaction potential, was then termed the ``Bouncing cDE
scenario",  
and has been shown to determine potentially observable effects on the
number counts of massive clusters as a function of redshift
\citep[][]{Baldi_2012b}. All the models included in the {\small CoDECS}  
suite have the same background cosmological parameters at $z=0$
consistent with the latest results from WMAP-7
\citep{komatsu_etal_2011}, summarised in 
Table~\ref{tab:parameters}, although the modified DE component makes
the dynamics of both the cosmic background and the cosmological
perturbations differ from $\Lambda$CDM. Linear density
perturbations are normalised to the same amplitude 
at the redshift of the last scattering surface $z_{\rm ls}\approx
1100$, and evolved forward in time with the specific growth factor
$D_{+}(z)$ of each model until the starting redshift of the  
simulations, $z_{\rm i}=99$. The growth factor was obtained by numerically integrating Eqs.~(\ref{gf_c})-(\ref{gf_b}) along the respective background solution. 
Initial conditions of all the {\small CoDECS} simulations have been produced by setting up a random-phase realisation of the linear matter power spectrum obtained by running the public 
Boltzmann code {\small CAMB}\footnote{www.cosmologist.info} for a
$\Lambda $CDM cosmology with the parameters of
Table~\ref{tab:parameters}, and rescaling particles displacements from
a {\em glass} distribution using the different growth factors as
described above. Therefore, all the simulations virtually share
exactly the same initial conditions at the last scattering surface. 

As anticipated above, for the present work we restrict our
investigation to a few selected models among the six cosmologies
available within the {\small CoDECS} Project, with the aim of
describing the main effects that 
cDE cosmologies generate on CMB lensing observables. Furthermore, the
{\small CoDECS} 
suite will be soon extended to a wider range of cDE cosmologies, which
will allow for an extensive application of our numerical approach in
the near future. Our selected sample of models includes 
the reference $\Lambda $CDM scenario, the most strongly coupled model
($\beta _{c}=0.15$) among those with a constant positive coupling plus
an exponential potential, labelled ``EXP003", and, finally, the 
Bouncing cDE model characterised by a constant negative coupling
($\beta _{c}=-0.15$) plus a SUGRA potential, labelled ``SUGRA003". 
 We note that EXP003 represents quite a natural choice, because Weyl scaling
  of a scalar-tensor theory leads to a roughly exponential potential
  with a constant coupling (as illustrated in detail, e.g., in
  \cite{Pettorino_Baccigalupi_2008}).
The SUGRA model represents an alternative model, somehow much more
peculiar - extreme scenario - due to the presence of the bouncing
(although the SUGRA potential can be predicted from theory). In this
way, by comparing the two models, we actually 
test both a simple coupled dark energy model and a more complex 
scenario, therefore providing a good hint of the 
effects that can be expected by such models compromised with the need
of feasibility for implementation in Boltzmann codes and N-Body
simulations. Furthermore, the choice of these models is also
motivated by the fact that these two 
cDE scenarios show very different background and linear perturbations
evolutions, 
despite having the same absolute value of the coupling $|\beta
_{c}|=0.15$, and therefore the same strength of the associated 
fifth force, given by the factor $\Gamma _{c}$ in Eq.~(\ref{gf_c})
which for both models takes the value $\Gamma _{c}=1.03$,
corresponding to a $3\%$ enhancement in the gravitational attraction  
for CDM particles. The different impact of these two models on
observable quantities associated to the formation of linear and
non-linear structures, including the CMB lensing signal, will 
then be entirely determined by the different dynamics of the DE scalar
field and by the related evolution of the CDM particle mass variation
and extra-friction term introduced in 
Eqs.~(\ref{mass_variation}) and (\ref{gf_c}), respectively.
 In this respect we note that, in order to have, for all the
  considered models, the same 
  cosmological parameters at $z=0$, while producing different
  perturbations dynamics, the SUGRA model requires some additional 
  fine-tuning with respect to the EXP003 model: $\alpha$ and $\beta$ parameters
  need to be chosen such that the scalar field $\phi$ goes at $z = 0$
  approximately in the same position it was at $z_{\rm CMB}$. 
The viability of these models in terms of CMB
observables has yet to be properly investigated, in particular for
what concerns the impact of variable-coupling and bouncing cDE models
on the large-scale power of CMB anisotropies. Although such analysis
might possibly lead to tighter bounds on the coupling and on the
potential functions than the ones allowed in the present work, here we
are mainly interested in understanding the impact of cDE scenarios on
the secondary anisotropies induced on CMB by large-scale structure
lenses at late times, with a particular focus on the role played by non-linear effects, and
we deliberately choose quite extreme values of the models parameters
in order to maximize the effects under investigation.

The relevant parameters of the cosmological models considered in this
work are summarised in Table~\ref{tab:models}.  
For further details about the simulations we refer the interested reader to the {\small CoDECS} paper \citep{CoDECS}. 
\begin{table}
\begin{center}
\begin{tabular}{cc}
\hline
Parameter & Value\\
\hline
$H_{0}$ & 70.3 km s$^{-1}$ Mpc$^{-1}$\\
$\Omega _{\rm CDM} $ & 0.226 \\
$\Omega _{\rm DE} $ & 0.729 \\
${\cal A}_{s}$ & $2.42 \times 10^{-9}$\\
$ \Omega _{b} $ & 0.0451 \\
$n_{s}$ & 0.966\\
\hline
\end{tabular}
\end{center}
\caption{ {\small CoDECS} cosmological parameters at $z=0$: consistently with
the notation adopted in the literature \citep{komatsu_etal_2011},
$H_{0}$ is the Hubble expansion rate at present, $\Omega_{x}$ the density ratio with respect to the
cosmological critical density of the species $x$; $A_{s}$ and $n_{s}$
represent the amplitude and spectral index of the initial power law
scalar perturbation spectrum, respectively.}
\label{tab:parameters}
\end{table}
\begin{table*}
\begin{center}
\setlength{\tabcolsep}{6pt}
\begin{tabular}{llccccccc}
Model & Potential  &  
$\alpha $ &
$\beta _{c}$ &
\begin{minipage}{45pt}
{\begin{center}
Scalar field \\ normalisation
\end{center}}
\end{minipage} &
$w_{\phi }(z=0)$ &
$\sigma _{8}(z=0)$\\
\\
\hline
$\Lambda $CDM & $V(\phi ) = A$ & -- & -- & -- & 
$-1.0$ & 
$0.809$ \\
EXP003 & $V(\phi ) = Ae^{-\alpha \phi }$  & 0.08 & 0.15 & $\phi (z=0) = 0$ & 
$-0.992$ & 
$0.967$\\
SUGRA003 & $V(\phi ) = A\phi ^{-\alpha }e^{\phi ^{2}/2}$  & 2.15 &
-0.15 & $\phi (z\rightarrow \infty ) = \sqrt{\alpha }$ &  
$-0.901$ & 
$0.806$ \\
\hline
\end{tabular}
\caption{Cosmological models from {\small CoDECS} which are considered in the
  present work and their specific parameters.}
\label{tab:models}
\end{center}
\end{table*}

\section{CMB lensing}
\label{sec:cmb_lensing}
CMB photons are deflected from an original direction ${\bf \hat{n}}'$ on the last scattering surface to a direction ${\bf \hat{n}}$ on the observed sky, so that the lensed CMB field is given by 
$\tilde{X}({\bf \hat{n}}) = X({\bf \hat{n}}')$ in terms of the unlensed field $X=T,Q,U$, representing the total intensity and Stokes parameters for linear polarisation, respectively. 
As in previous works \citep{carbone_etal_2008,carbone_etal_2009}, we consider here the case in which the change in the comoving separation of CMB photon trajectories induced by lensing is small 
compared to the comoving separation of the undeflected rays. In this case it is sufficient to calculate the relevant quantities for lensing, including the deflection angle, along the undeflected 
rays. This approach corresponds to the Born approximation. 
Adopting conformal time and comoving coordinates in a flat Friedmann
Robertson Walker (FRW) geometry, the integral for the
projected (along the line of sight) lensing-potential due to scalar perturbations with no anisotropic stress reads 
\begin{align}
\label{lensingpotential}
\Psi({\bf \hat{n}})\equiv 
-2\int_0^{r_*} \frac{r_*-r}{r_*r}\,\frac{\Phi(r{\bf\hat{n}};\eta_0-r )}{c^2}\,{\rm d}r\,,
\end{align}
while the corresponding deflection angle integral is
\begin{align}
\label{deflection angle}
\boldsymbol{\alpha}({\bf \hat{n}})\equiv 
-2\int_0^{r_*}
\frac{r_*-r}{r_*r}\,\nabla_{\bf\hat{n}}\frac{\Phi(r{\bf\hat{n}};\eta_0-r )}{c^2}\, {\rm d}r\,,
\end{align}
where $r$ is the comoving distance, $r_*\simeq 10^4$ Mpc is its value at the last-scattering surface, $\eta_0$ is the present conformal time, $\Phi$ is the gravitational potential
generated by density perturbations, and $[1/r]\nabla_{\hat{\bf n}}$ is the two dimensional (2D) transverse derivative with respect to the line of sight pointing in the direction 
${\hat{\bf n}}\equiv(\vartheta,\varphi)$. The vector ${\bf \hat{n}}'$ is obtained from ${\bf \hat{n}}$ by moving its end on the surface of a unit sphere by a distance 
$|\nabla_{\bf\hat{n}}\Psi({\bf \hat{n}})|$ along a geodesics in the direction of $\nabla_{\bf\hat{n}}\Psi({\bf \hat{n}})$. We assume $|\nabla_{\bf\hat{n}}\Psi({\bf \hat{n}})|$ 
to be constant between ${\bf \hat{n}}$ and ${\bf \hat{n}}'$,
consistently with the Born approximation. The latter has been shown to
hold down to the arcminute scale, taking into 
account the non-linearities corresponding to such scales
\citep{hirata_seljak_2003,shapiro_cooray_2006}. If the gravitational
potential $\Phi$ is Gaussian, the lensing potential is also Gaussian. 
However, the lensed CMB is non-Gaussian, as it is a second order cosmological effect produced by cosmological perturbations onto CMB anisotropies, yielding a finite correlation 
between different scales and thus non-Gaussianity. This is expected to
be particularly important especially at small scales, i.e. tens of comoving Mpc or less, 
due to the the fact that non-linearities are already present in the
underlying properties of lenses. It is typically more accurate to
solve the integral in Eq.~(\ref{deflection angle}) directly to obtain
the deflection angle instead of finite differencing the lensing
potential \citep{lewis_challinor_2006,bartelmann_schneider_2001}.
Semi-analytical techniques, such as {\small LensPix} and {\small
lenS2HAT} \citep{lewis_2005, fabbian_stompor2013},  exist for calculating Gaussian realisations of the lensing deflection angle, and therefore lensed
CMB maps, starting from the lensing potential power spectrum, including non-linear corrections on small scales \citep{smith_etal_2003}. This approach has been used to study
and validate CMB lensing templates obtained via ray tracing through
the Millennium Run N-body simulation at the level of the two point
correlation function  
\citep{carbone_etal_2009}, finding significant deviations from the
semi-analytical expectations at multipoles of order $10^{3}$. On large
scales, corresponding to about 1/3 of a degree in the sky or more, due to the limited box-side
size, $L= 500$ Mpc/$h$ comoving, of the Millennium Run, the 
lensing signal was not correctly reproduced. Therefore, \citet{carbone_etal_2008,carbone_etal_2009}, as
a feasibility study, imported synthetic maps of lensing potential,
deflection angle and lensed CMB generated through {\small Lenspix},
filtered on large scales, in order to achieve the complete angular
power spectrum (for further details see \citet{carbone_etal_2008,carbone_etal_2009}). Even if the CoDECS  
simulations still represent only a portion of the Hubble volume, this
technique is not applied in the present analysis since we are
interested to the pure N-body signal and its response to the
underlying DE behaviour. In the next two sub-sections we outline our
ray-tracing and sky projection procedures.  

\subsection{Ray tracing}
\label{sec:rt}
\begin{figure*}
\begin{center}
\includegraphics[width=0.7\textwidth]{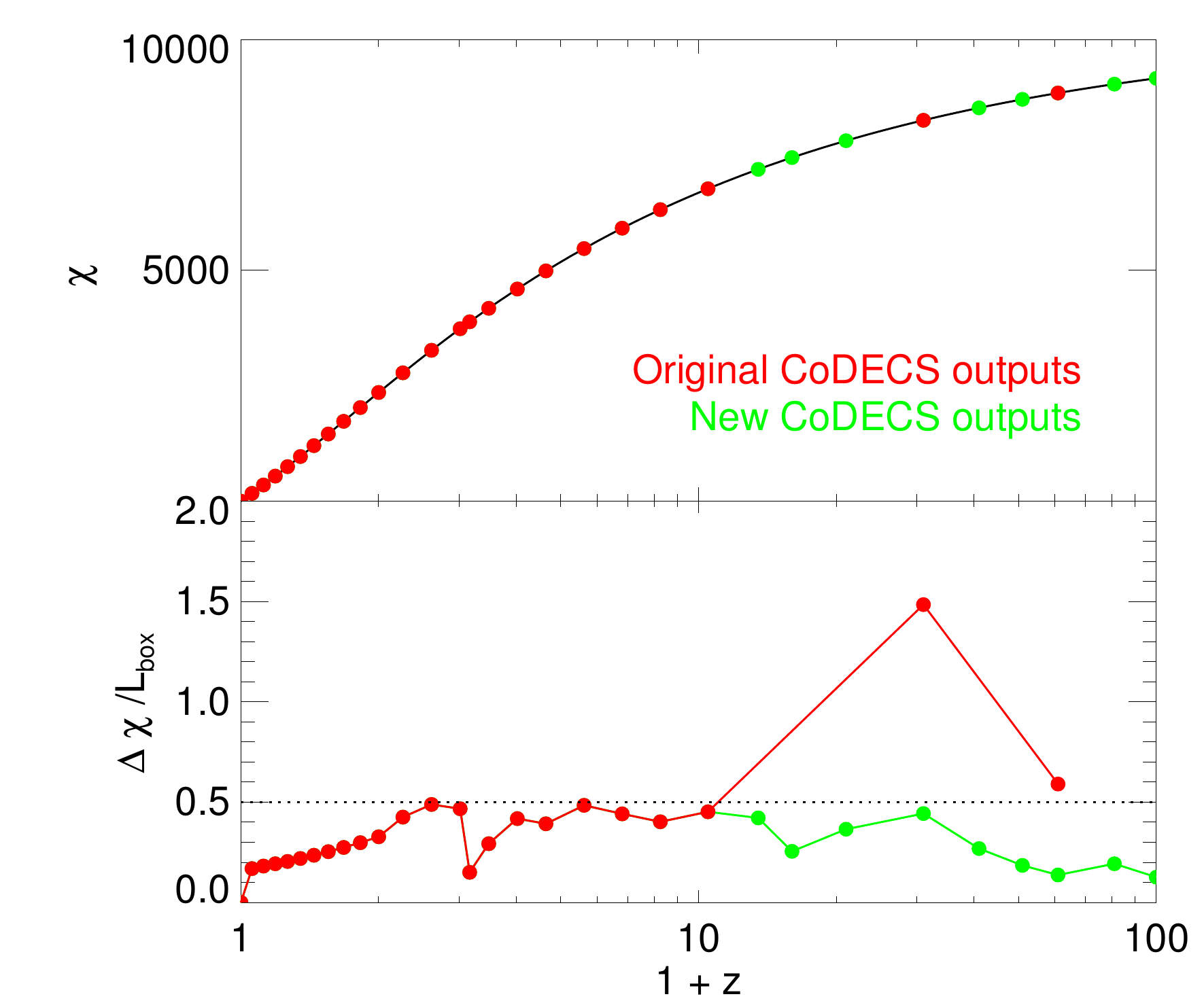}
\caption{Top: comoving distance of the gravitational potential snapshots
  extracted from the CoDECS simulations. Bottom: 
ratio of the comoving spacing with respect to the box side. The red
data correspond to the original spacing in 
CoDECS, which was re-computed to become the one corresponding to the
green points, in order to implement the CMB ray-tracing across the
gravitational potential of the simulations.}
\label{fig:snapshots}
\end{center}
\end{figure*}

The lensing potential maps obtained through ray-tracing across the
CoDECS structures have been generated as described below.\\
First, we pre-compute and store the gravitational 
potential grids from the simulation snapshots, taking care of achieving a dense
sampling in redshifts. In Fig.~\ref{fig:snapshots}, we show the grid
spacing $\chi$ representing the comoving distances of the snapshots as
a function of the redshift, together with their
ratio to the {\small CoDECS} box-side size, $L_{\rm box}=1$ Gpc$/h$. Notice that we had to re-sample the 
simulation outputs with respect to the public version, in order to
achieve a proper sampling in $z$ for CMB 
lensing analyses. The original redshift sampling of the CoDECS
snapshots is shown in 
red in the upper and lower panels of Fig.~\ref{fig:snapshots}.\\
Second, in order to produce mock lensing potential maps that cover the
past light cone over the full sky, we stack the gravitational
pre-computed potential grids around the observer following the
technique developed in \citep{carbone_etal_2008}.\\
Third, the total volume around the observer up to $z_{*}$ is divided into spherical shells, each one with a thickness corresponding to $L_{\rm box}$. All the boxes falling into the same 
shell are translated and rotated with the same random vectors
generating a homogeneous coordinate transformation throughout the
shell, while the randomisation vectors change from shell to shell \citep{carbone_etal_2008}.\\
Finally, the peculiar gravitational potential at each point along a
ray with direction ${\bf\hat{n}}$ is spatially interpolated from the
CoDECS grid which possesses a spatial resolution  
of $L_{\rm box}/2560\approx 400$ kpc$/h$. The deflection angle is
computed along the line of sight as well, by numerically evaluating the gravitational potential gradient and interpolating at each point along the line 
of sight. By varying the direction of
the light-ray along which the integration is performed, we construct
all sky maps of the lensing potential.

\subsection{Sky projection}
\label{sec:sp}
\begin{figure*}
\begin{center}
\setlength{\tabcolsep}{1pt}
\begin{tabular}{ll}
\resizebox{7.80cm}{!}{\includegraphics{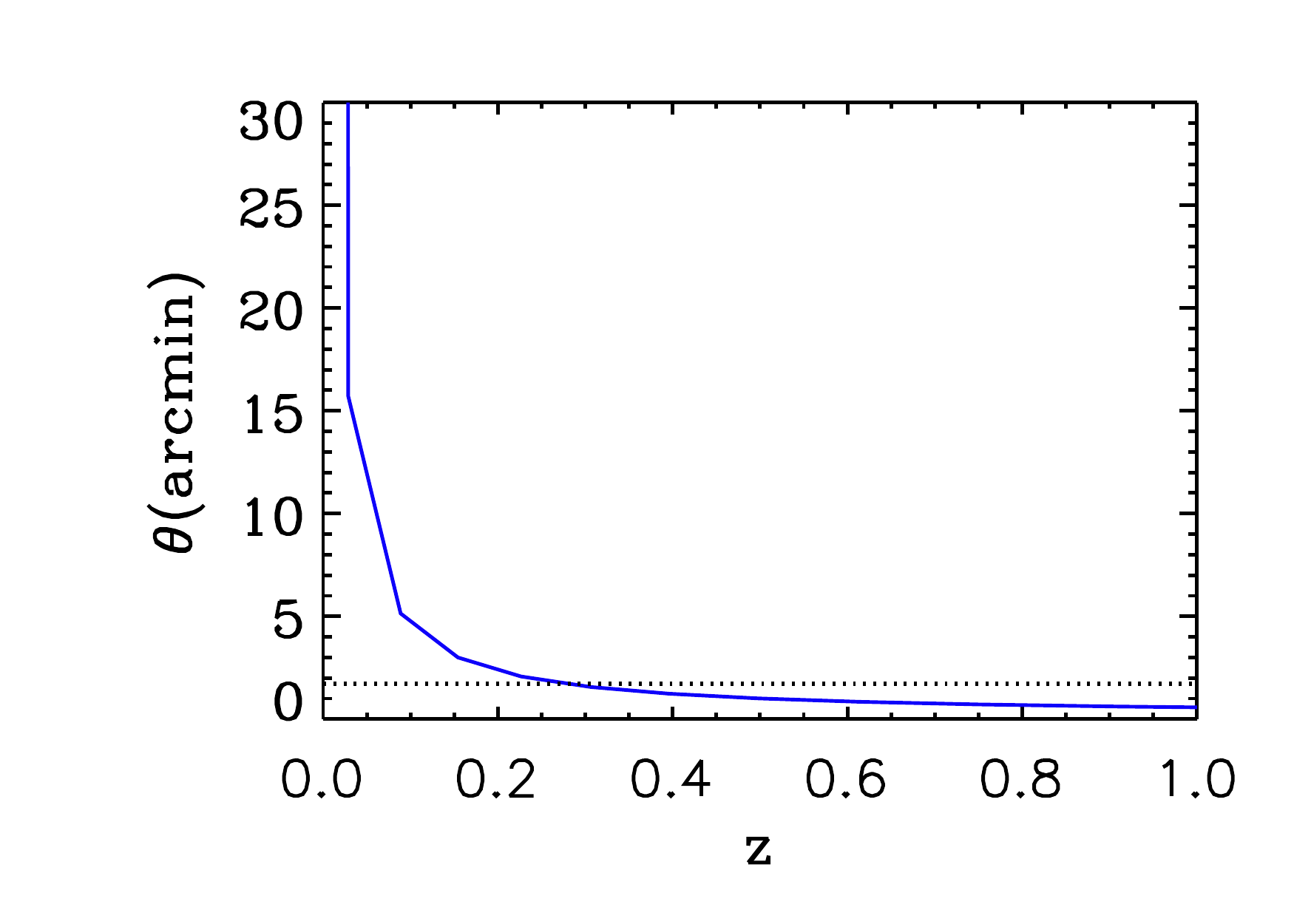}}
\resizebox{7.80cm}{!}{\includegraphics{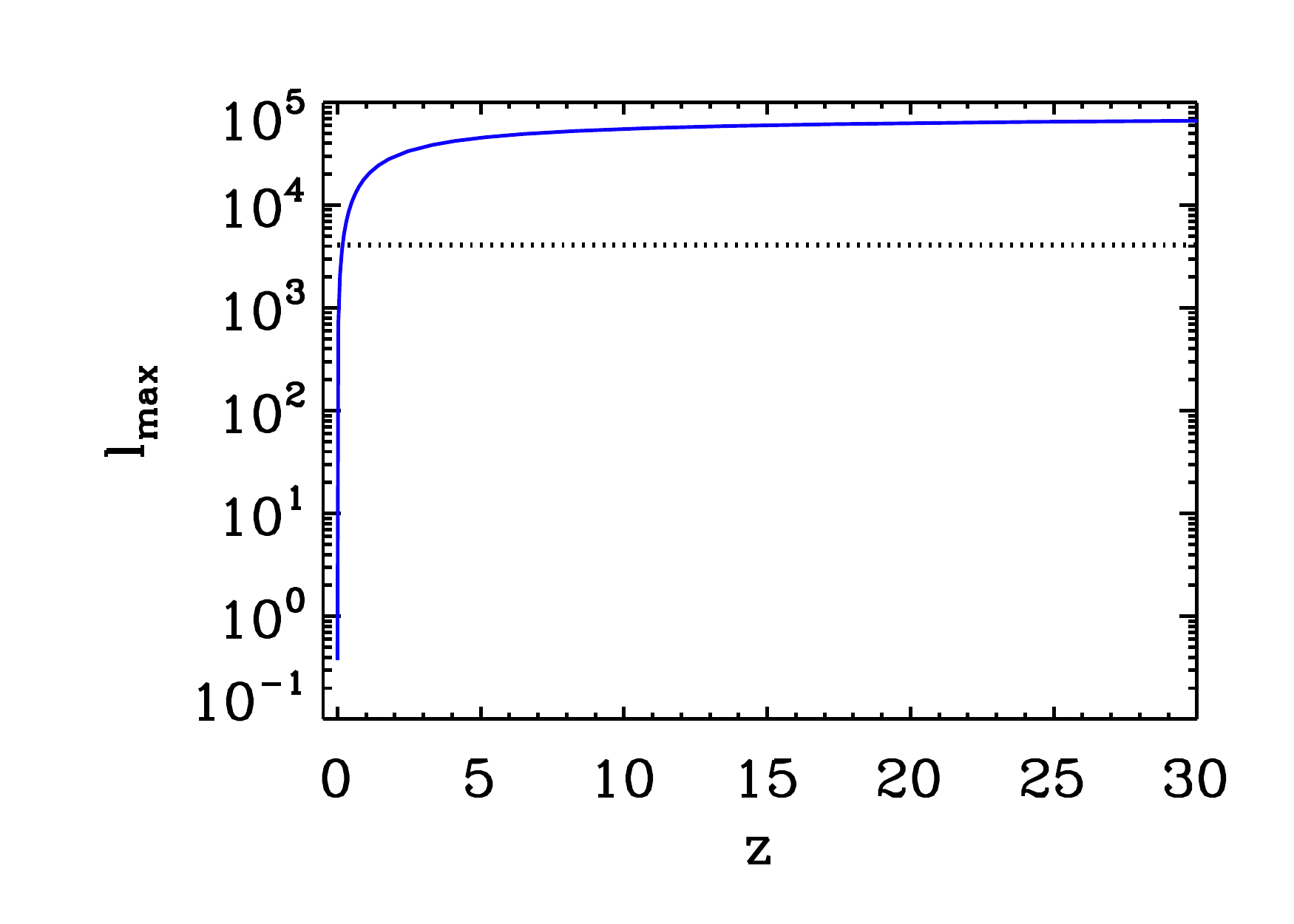}}
\end{tabular}
\caption{Left: comparison between the pixel size $\theta=1.72^{\prime}$ 
of the lensing-potential sky maps (dotted black line) and the redshift
dependent angular resolution  
(solid blue line) corresponding to the intrinsic spacing ($\sim 400 {\rm Mpc}/h$) of the
three-dimensional gravitational potential grids extracted from
CoDECS. Right: comparison between  
the maximum multipole $l_{max}$ corresponding to $2 \cdot N_{\rm side}$
and the redshift dependent maximum  
multipole $l_{max}(z)=180/\theta$ corresponding to the resolution of the potential grids.}
\label{fig:resolution}
\end{center}
\end{figure*}

We adopt the Hierarchical Equalised Altitute Pixel
(HEALpix\footnote{http://healpix.jpl.nasa.gov}) scheme for the
pixelisation of the sphere \citep{gorski_etal_2005}, and choose a
resolution of about $1.72$ arcminutes, corresponding to a resolution
parameter $N_{\rm side}=2048$; in an HEALPix environment, the latter
choice corresponds to angular  
multipoles of at least $l_{max}\simeq 2\cdot N_{\rm nside}=4106$,
before pixel effects start to kick in. The chosen angular resolution
has been checked as follows.  
In the left panel of Fig.~\ref{fig:resolution} we compare the map
angular resolution of $1.72'$ (solid black line) with the effective
angular resolution $\theta(z)$  
corresponding to the intrinsic grid spacing ($\approx 400$ kpc/$h$) of
the three dimensional gravitational potential field as a function of
redshift. We observe that at redshifts $z<0.3$ the chosen angular
resolution of $1.72'$ becomes smaller than $\theta(z)$. However, since
most of the lensing signal  
comes from higher redshifts, the latter effect is negligible
\citep{carbone_etal_2008}. On the right panel of
  Fig.~\ref{fig:resolution}, we make a similar 
comparison, considering the maximum multipole, $l_{max}(z)=180/\theta(z)$, that can be exploited
given the intrinsic grid resolution. For most of the redshifts, the
latter is larger than $l_{max}=4106$. Nonetheless, we adopt the
conservative choice of limiting our analysis to $l<3000$ where Poisson
noise from low-redshift potential integration starts to dominate the
simulated CMB lensing signals.

\section{Results}
\label{sec:results}
\begin{figure*}
\begin{center}
\resizebox{8.2cm}{!}{\includegraphics{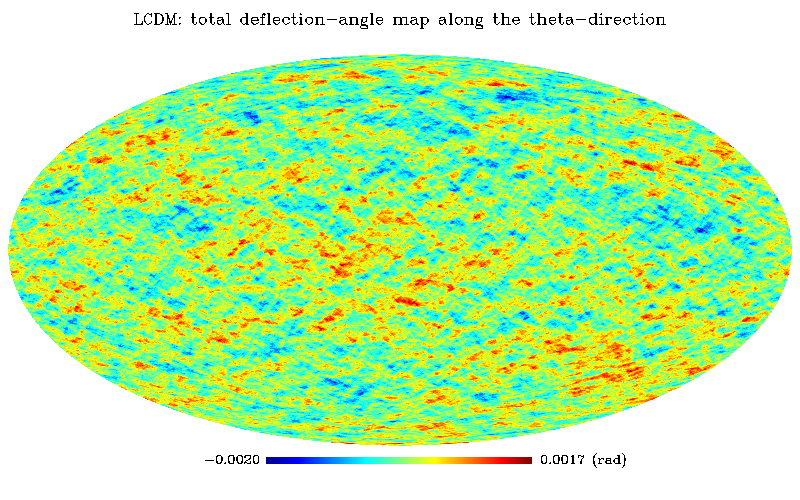}}
\resizebox{8.2cm}{!}{\includegraphics{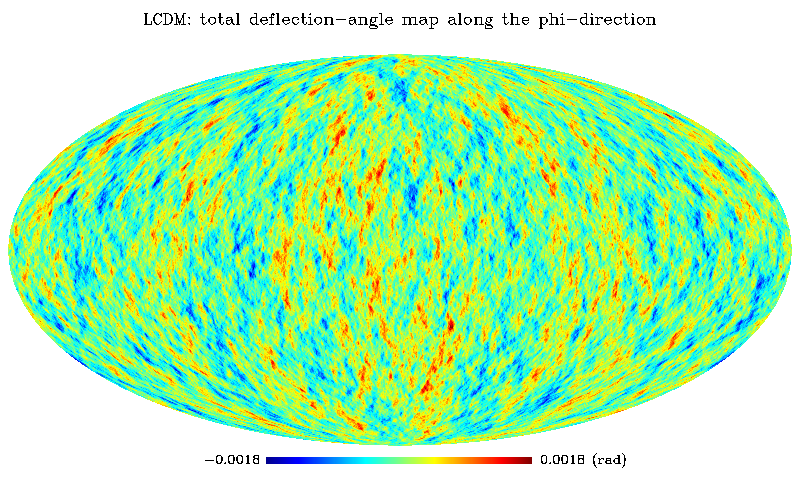}}
\resizebox{8.2cm}{!}{\includegraphics{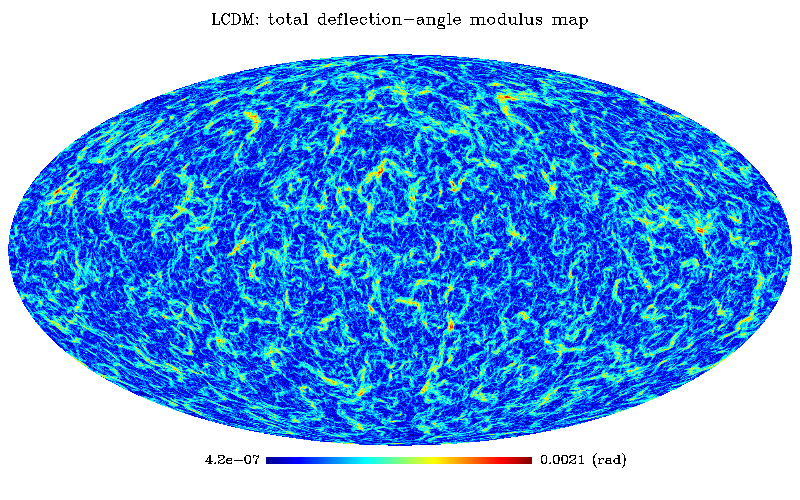}}
\resizebox{8.2cm}{!}{\includegraphics{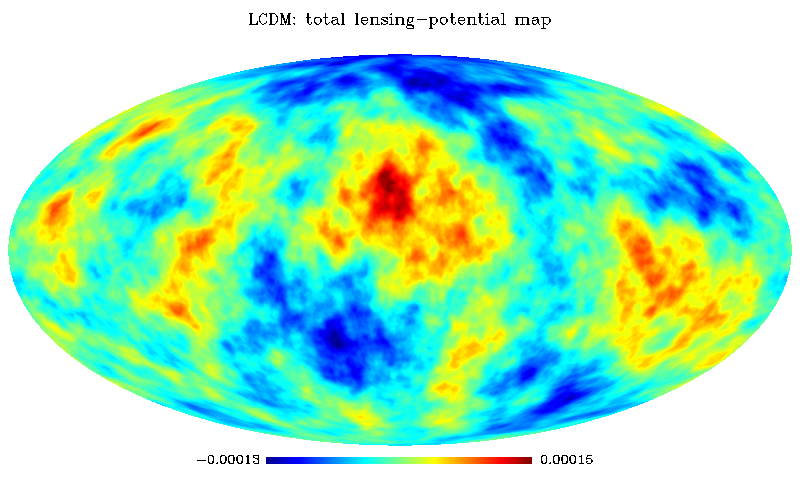}}
\caption{Maps of the relevant quantities for CMB weak-lensing for the
  $\Lambda$CDM model: deflection-angle components along the azimuth and elevation 
  angles (top panel: left and right, respectively), its magnitude (in radians) and
  the dimensionless lensing potential (bottom panel: left and right, respectively).} 
\label{fig:lcdmmaps}
\end{center}
\end{figure*}
\begin{figure*}
\begin{center}
\resizebox{8.2cm}{!}{\includegraphics{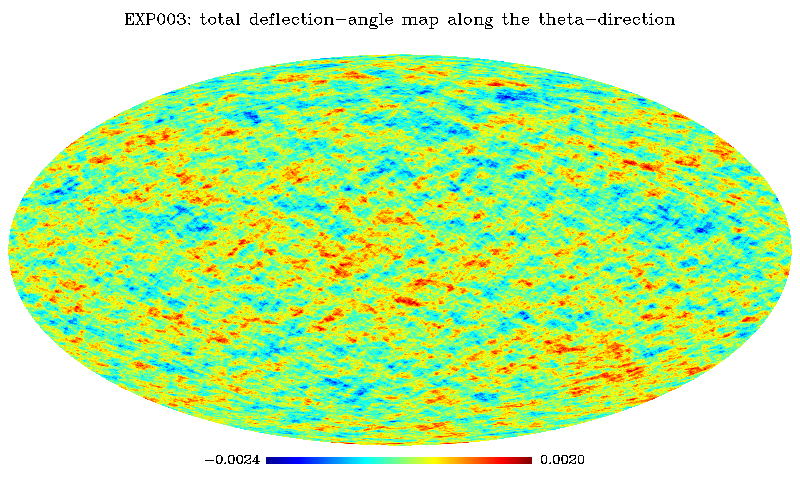}}
\resizebox{8.2cm}{!}{\includegraphics{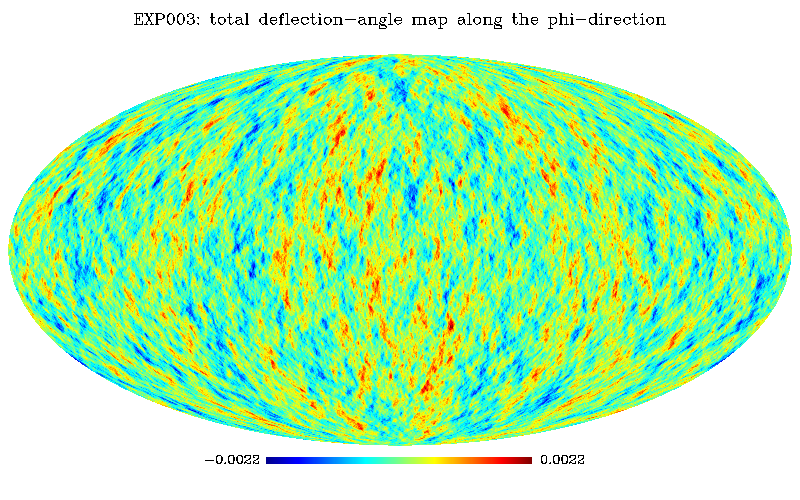}}
\resizebox{8.2cm}{!}{\includegraphics{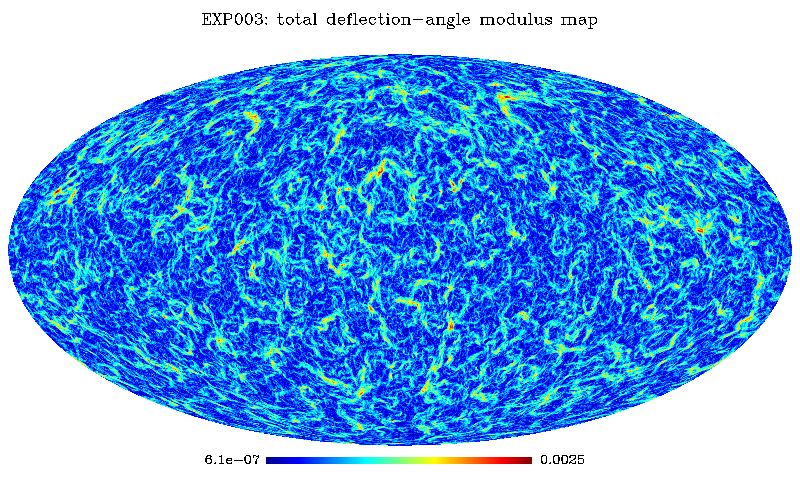}}
\resizebox{8.2cm}{!}{\includegraphics{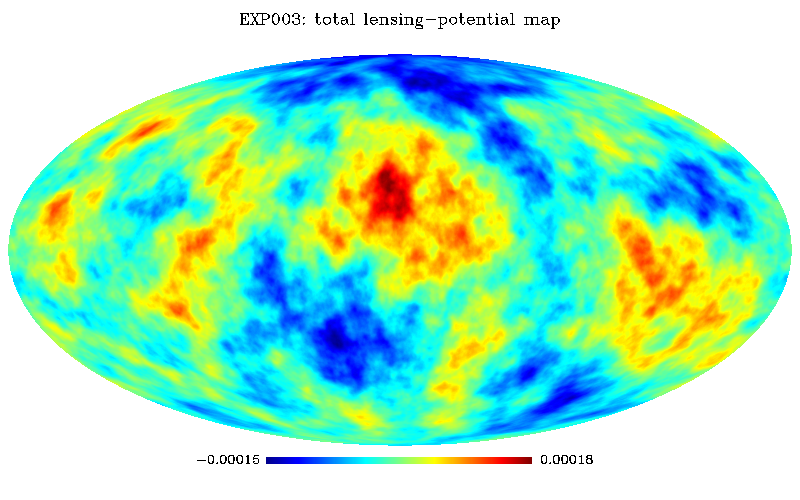}}
\caption{As in Fig.~\ref{fig:lcdmmaps} for the EXP003 model.}
\label{fig:exp003maps}
\end{center}
\end{figure*}
\begin{figure*}
\begin{center}
\resizebox{8.2cm}{!}{\includegraphics{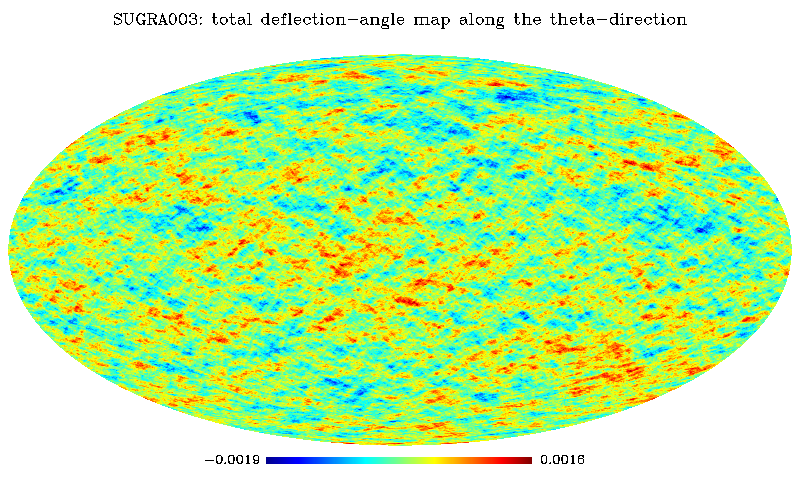}}
\resizebox{8.2cm}{!}{\includegraphics{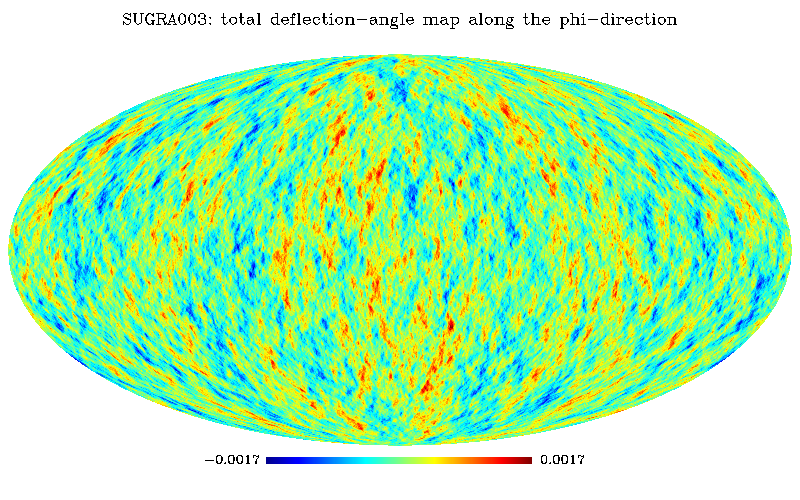}}
\resizebox{8.2cm}{!}{\includegraphics{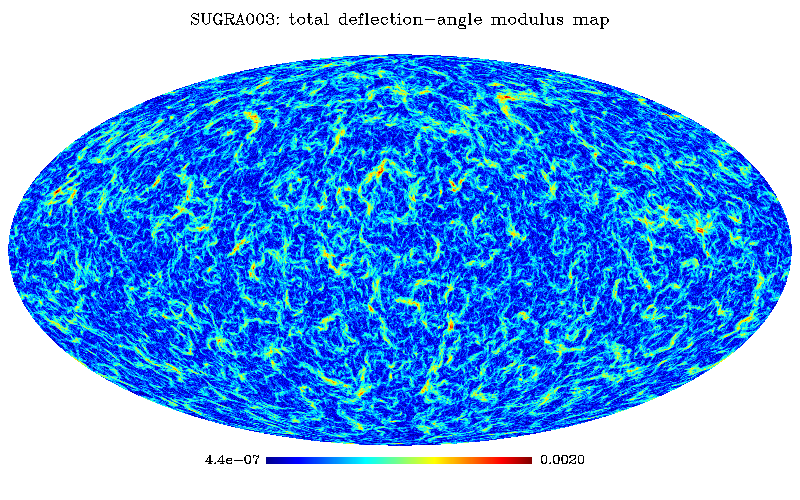}}
\resizebox{8.2cm}{!}{\includegraphics{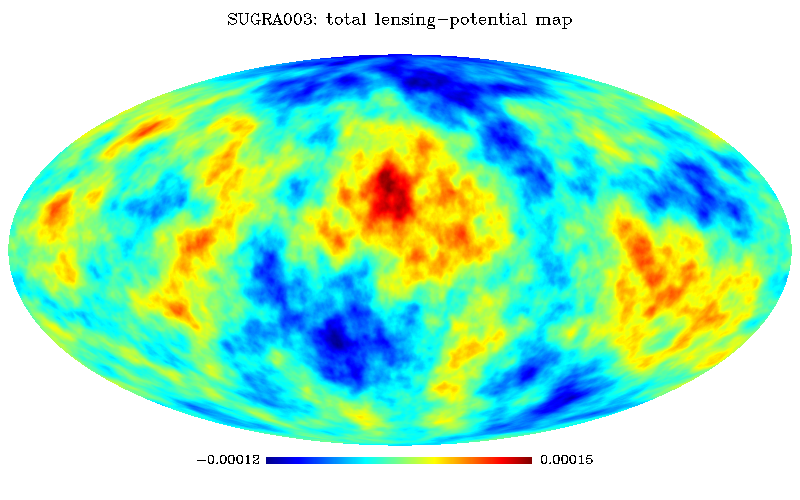}}
\caption{As in Figs.~\ref{fig:lcdmmaps}-\ref{fig:exp003maps} for the
  SUGRA003 model.} 
\label{fig:sugra003maps}
\end{center}
\end{figure*}
We now proceed in the illustration of our results. In
Figs.~\ref{fig:lcdmmaps}, \ref{fig:exp003maps}, and
\ref{fig:sugra003maps} 
we show the two components of the deflection angle
field ${\bf \alpha}$ along the $\theta$ and $\phi$ HEALPix spherical
coordinates 
(azimuth and elevation angles, respectively), its magnitude $|{\bf
  \alpha}|$, as well as the
projected lensing potential, for the three models described in
\S \ref{sec:codecs}, namely $\Lambda$CDM, EXP003, SUGRA003, respectively.
\begin{figure*}
\begin{center}
\resizebox{8.2cm}{!}{\includegraphics{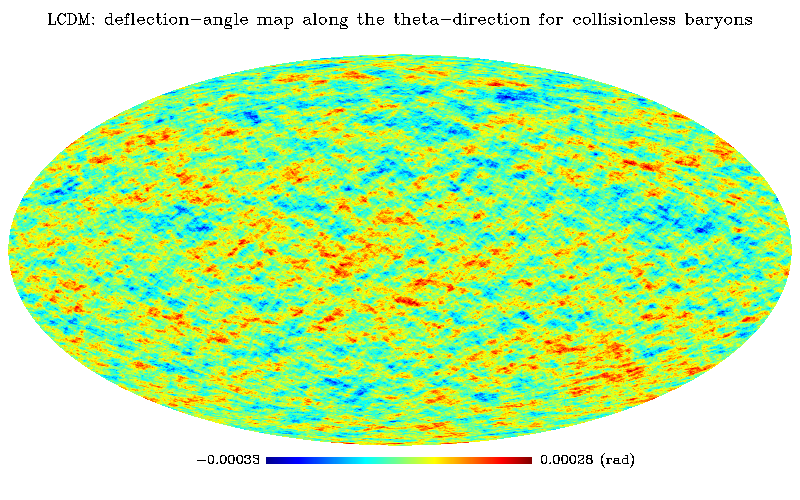}}
\resizebox{8.2cm}{!}{\includegraphics{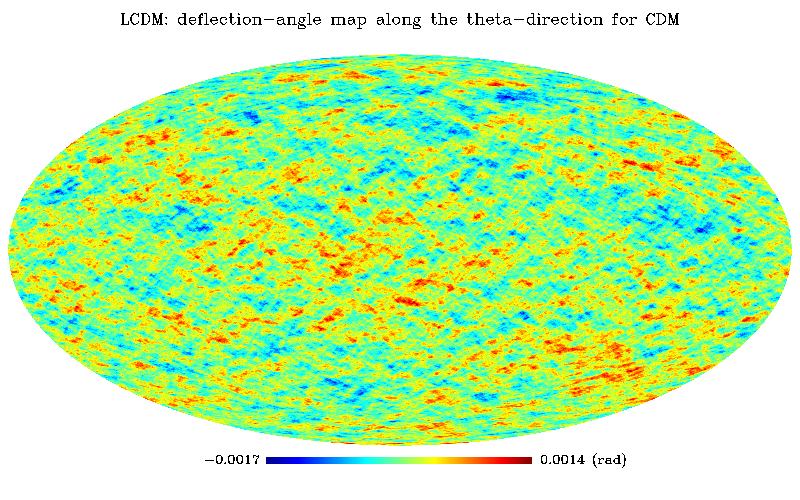}}
\resizebox{8.2cm}{!}{\includegraphics{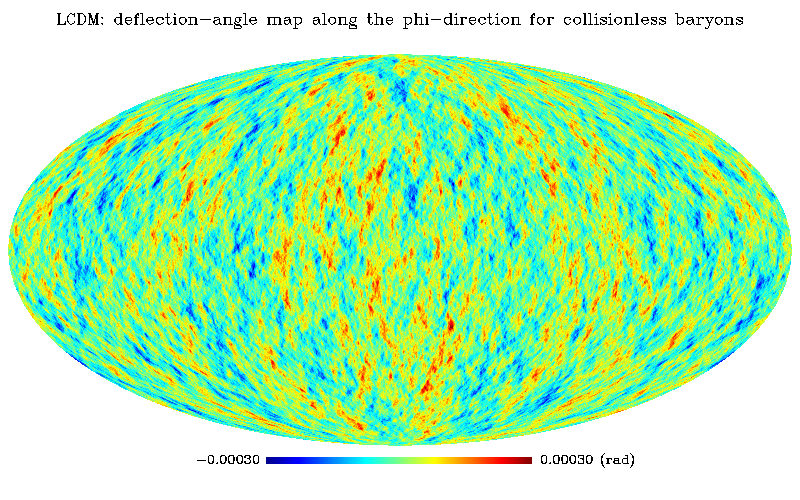}}
\resizebox{8.2cm}{!}{\includegraphics{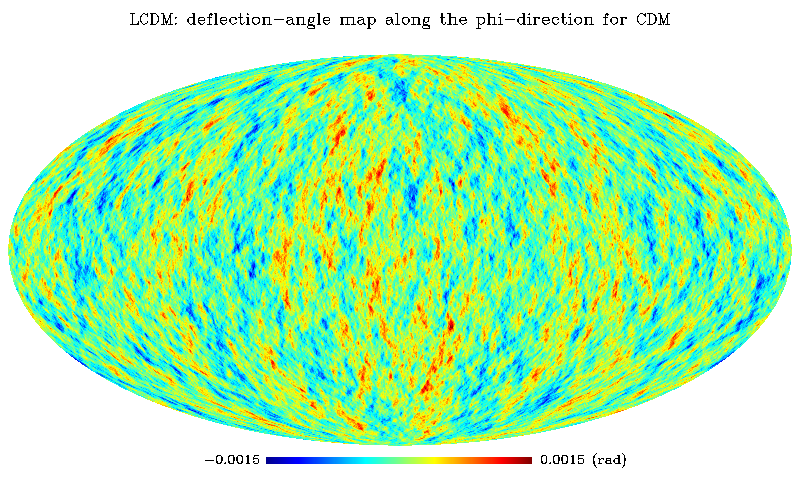}}
\resizebox{8.2cm}{!}{\includegraphics{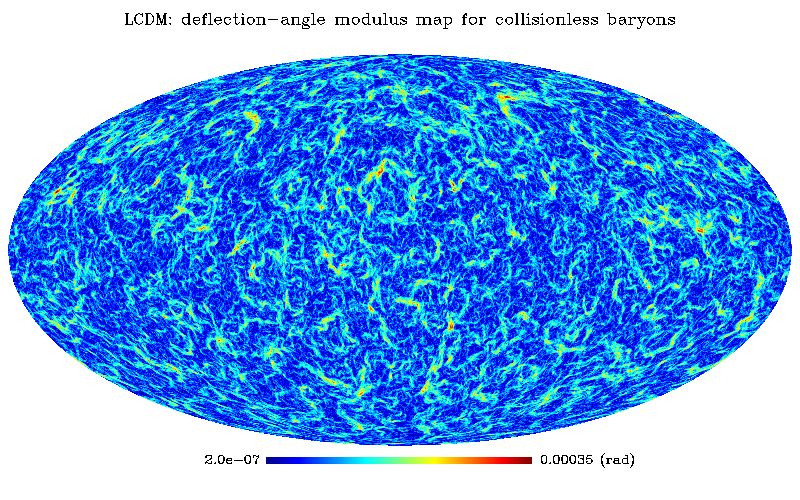}}
\resizebox{8.2cm}{!}{\includegraphics{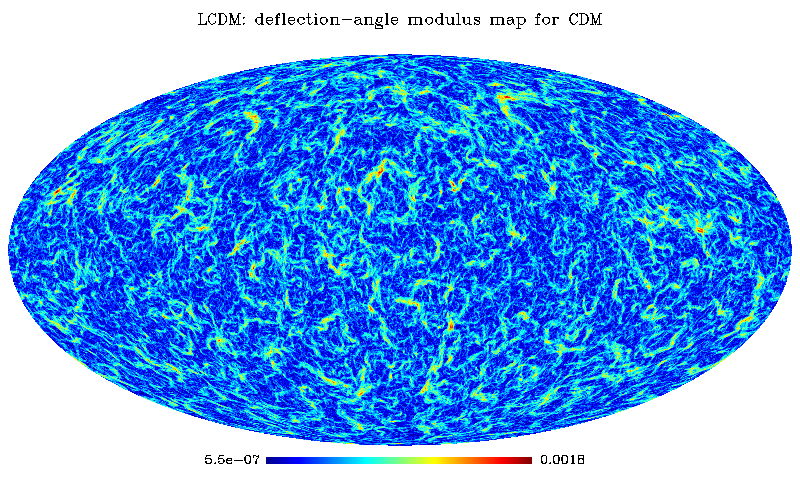}}
\resizebox{8.2cm}{!}{\includegraphics{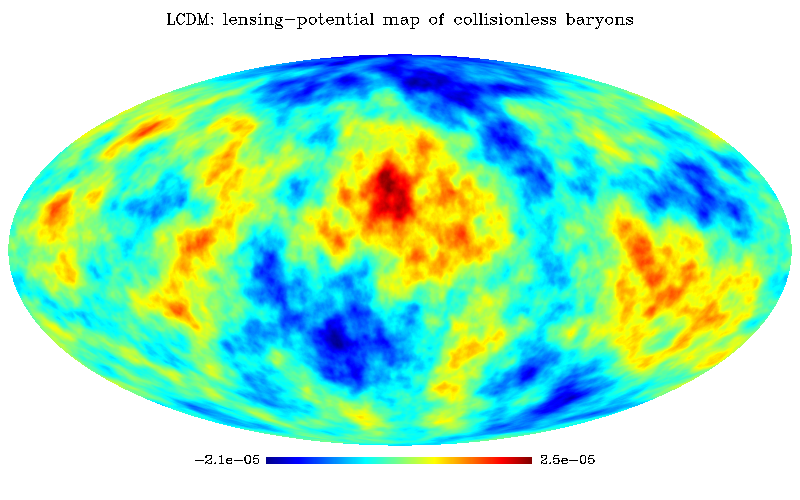}}
\resizebox{8.2cm}{!}{\includegraphics{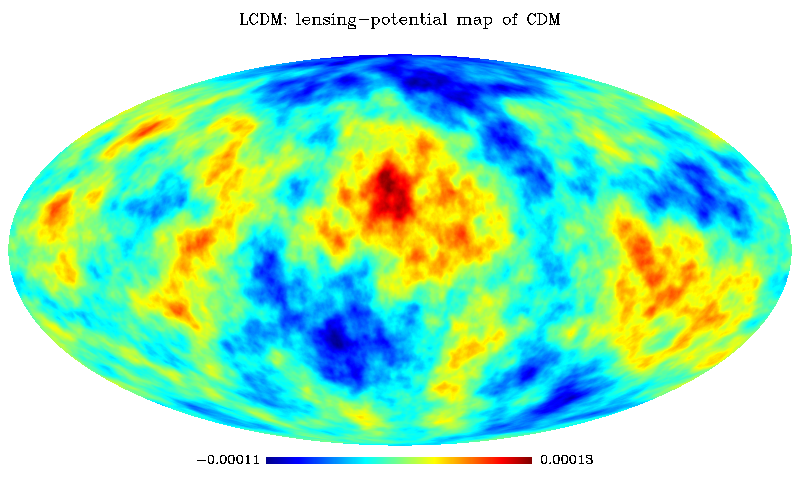}}
\caption{Maps of the relevant quantities for CMB weak-lensing in the
  $\Lambda$CDM case: from top to bottom,
  deflection angle components along the azimuth and elevation angles,
  magnitude and lensing potential (in 
  radians and dimensionless, respectively), for baryons (left) and CDM (right).}
\label{fig:lcdmmapsbcdm}
\end{center}
\end{figure*}
\begin{figure*}
\begin{center}
\resizebox{8.2cm}{!}{\includegraphics{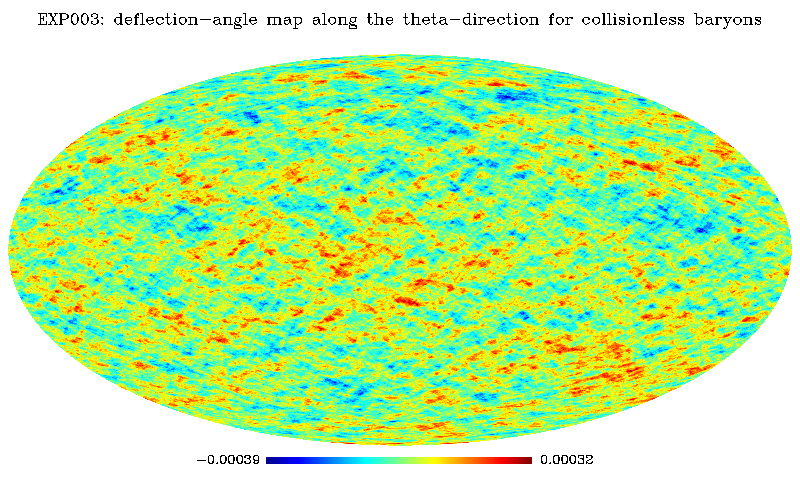}}
\resizebox{8.2cm}{!}{\includegraphics{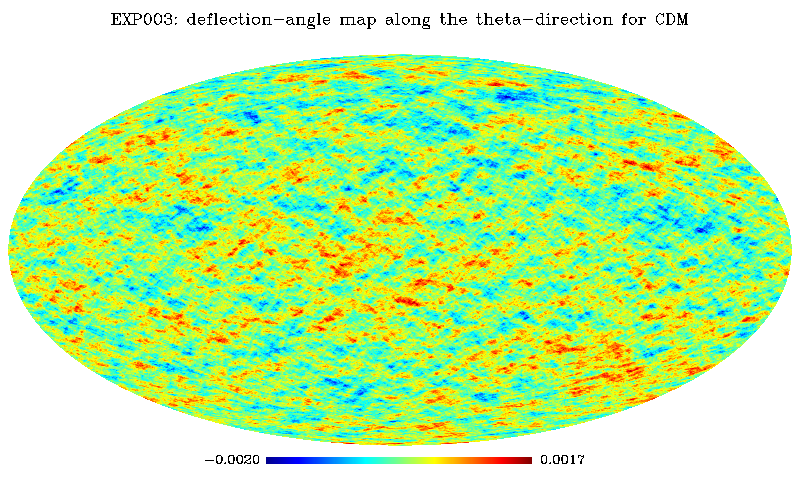}}
\resizebox{8.2cm}{!}{\includegraphics{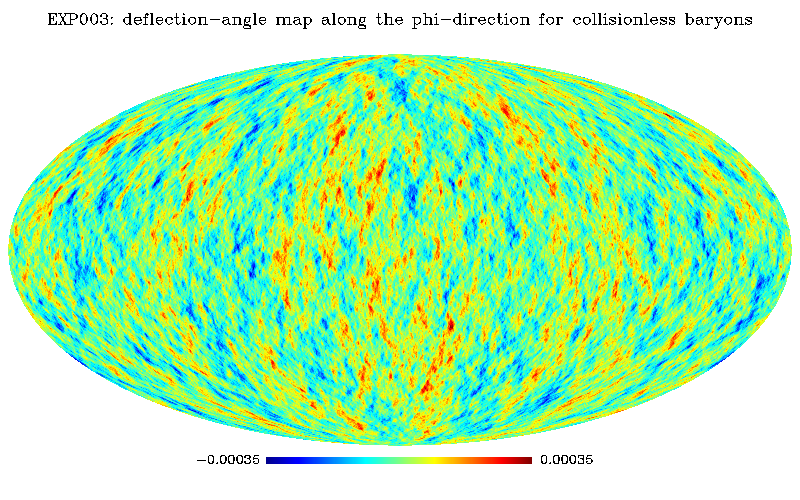}}
\resizebox{8.2cm}{!}{\includegraphics{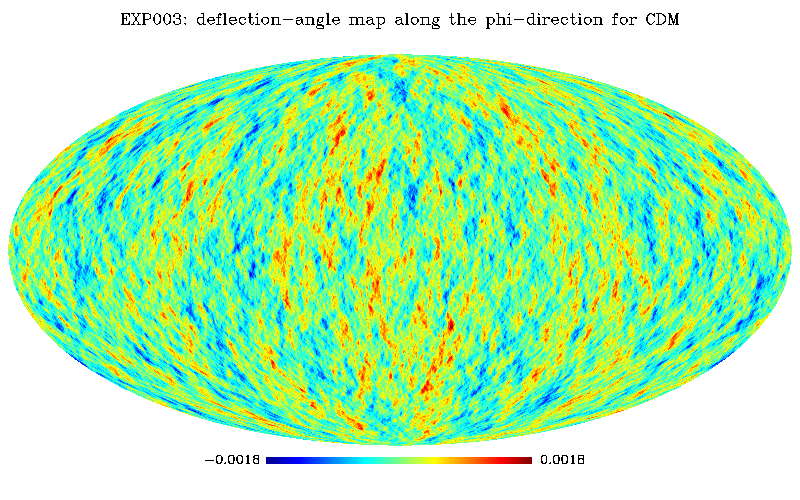}}
\resizebox{8.2cm}{!}{\includegraphics{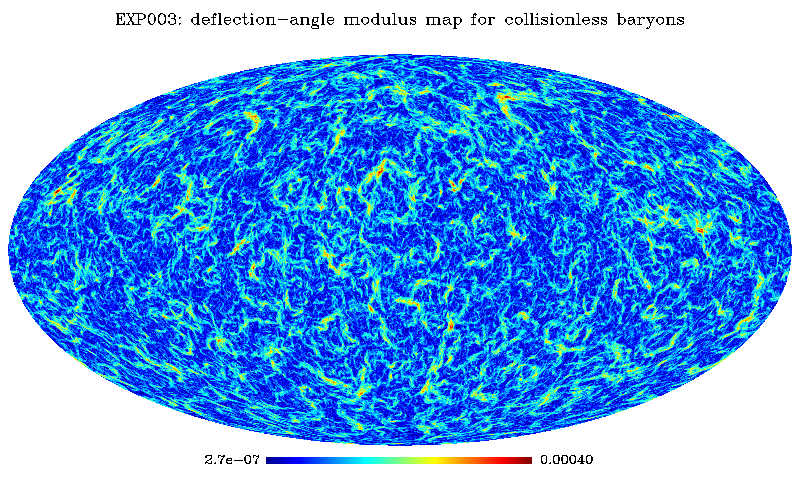}}
\resizebox{8.2cm}{!}{\includegraphics{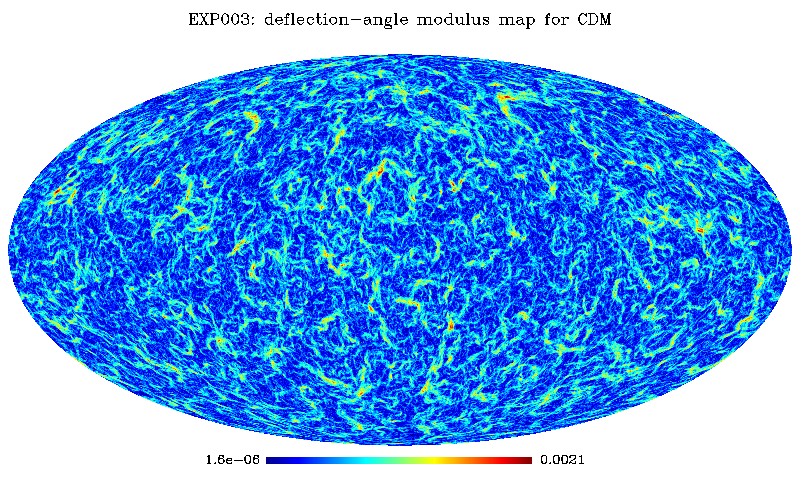}}
\resizebox{8.2cm}{!}{\includegraphics{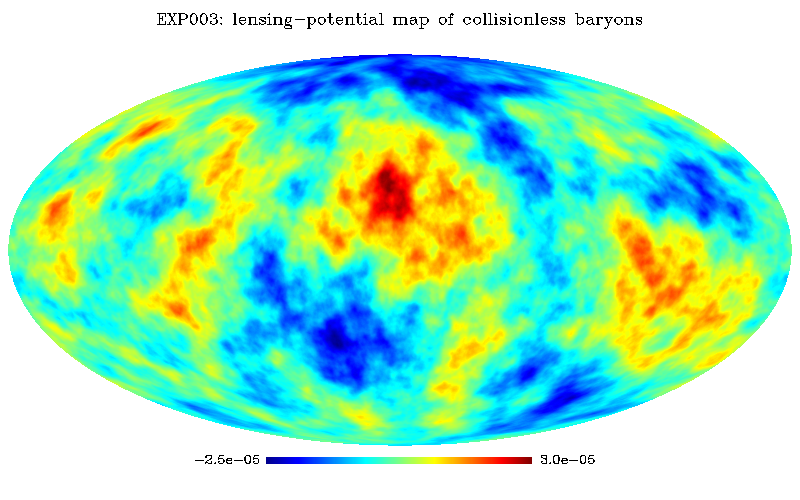}}
\resizebox{8.2cm}{!}{\includegraphics{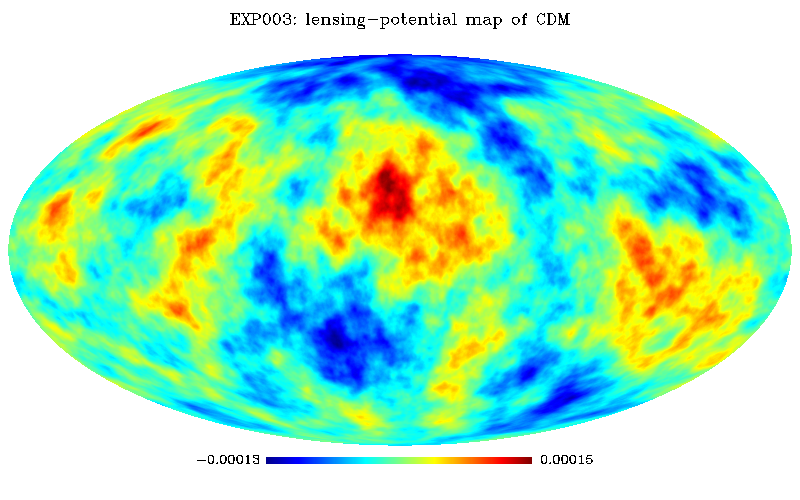}}
\caption{As in Fig.~\ref{fig:lcdmmapsbcdm} for the EXP003 model.}
\label{fig:exp003mapsbcdm}
\end{center}
\end{figure*}
\begin{figure*}
\begin{center}
\resizebox{8.2cm}{!}{\includegraphics{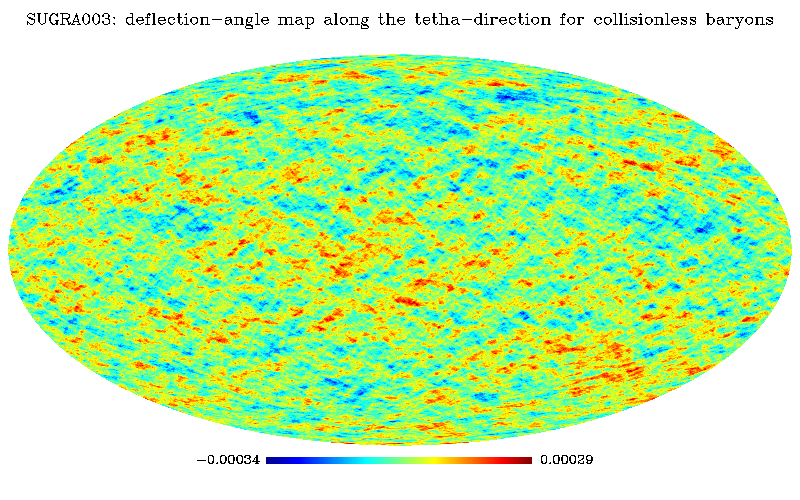}}
\resizebox{8.2cm}{!}{\includegraphics{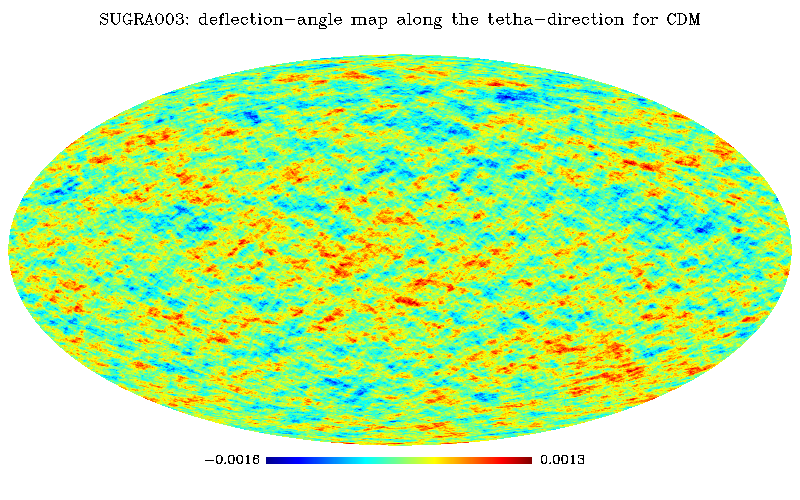}}
\resizebox{8.2cm}{!}{\includegraphics{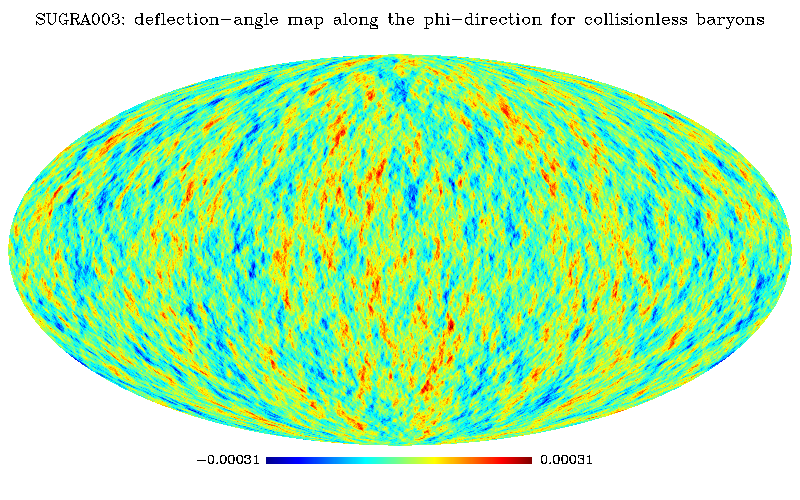}}
\resizebox{8.2cm}{!}{\includegraphics{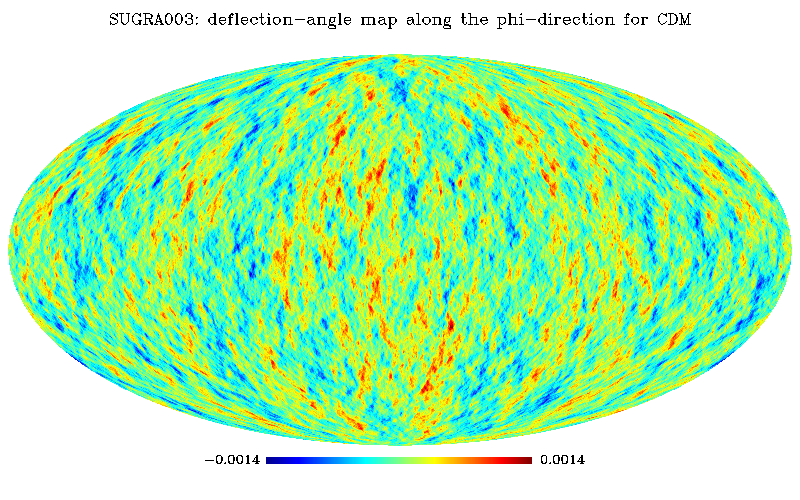}}
\resizebox{8.2cm}{!}{\includegraphics{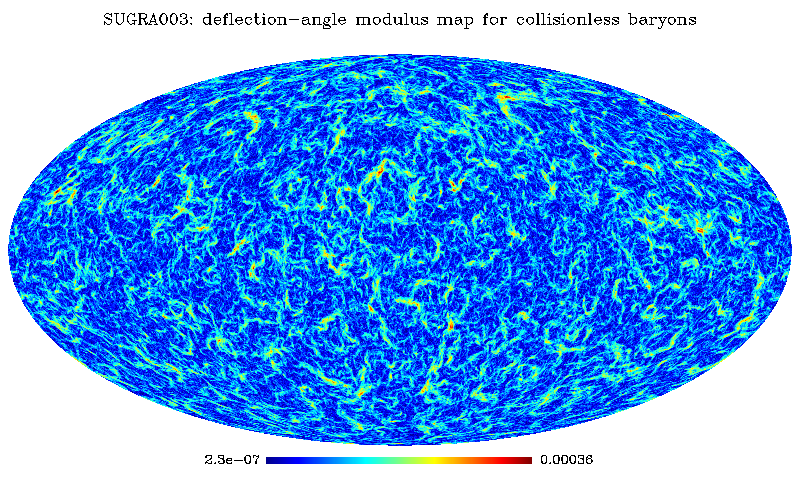}}
\resizebox{8.2cm}{!}{\includegraphics{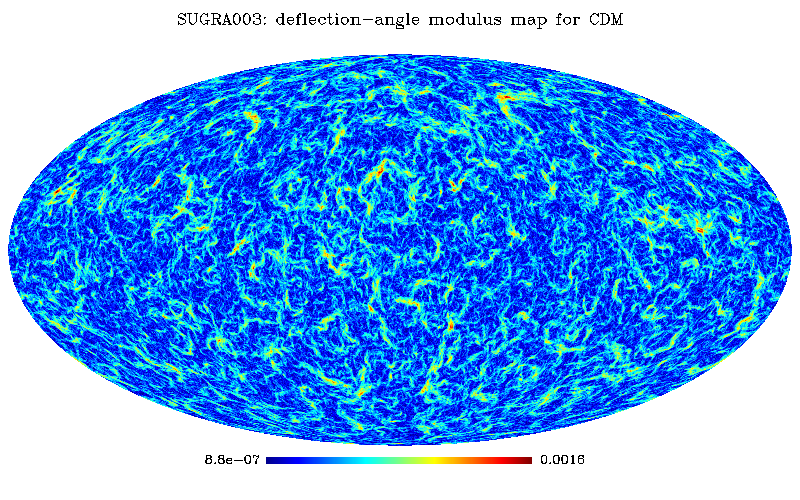}}
\resizebox{8.2cm}{!}{\includegraphics{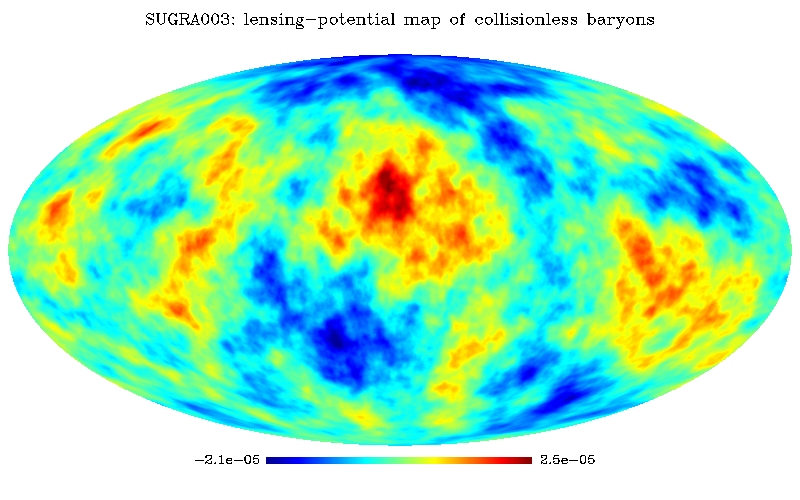}}
\resizebox{8.2cm}{!}{\includegraphics{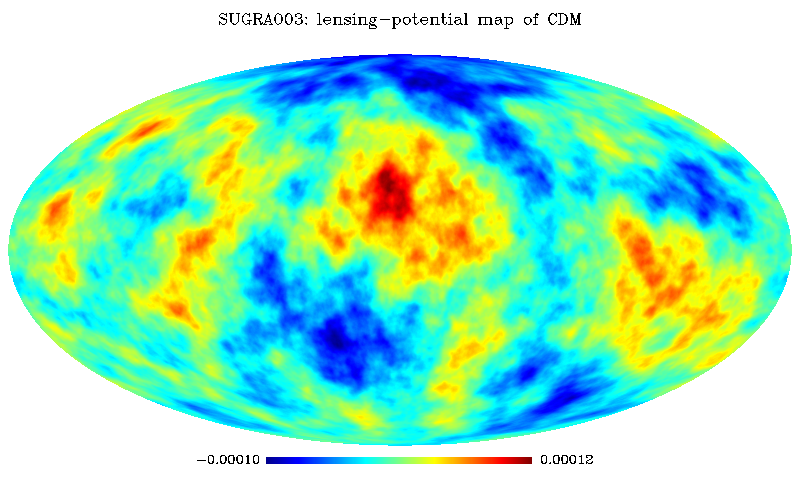}}
\caption{As in Figs.~\ref{fig:lcdmmapsbcdm}-\ref{fig:exp003mapsbcdm}
  for the SUGRA003 model.}
\label{fig:sugra003mapsbcdm}
\end{center}
\end{figure*}
\begin{figure*}
\begin{center}
\includegraphics[width=0.8\textwidth]{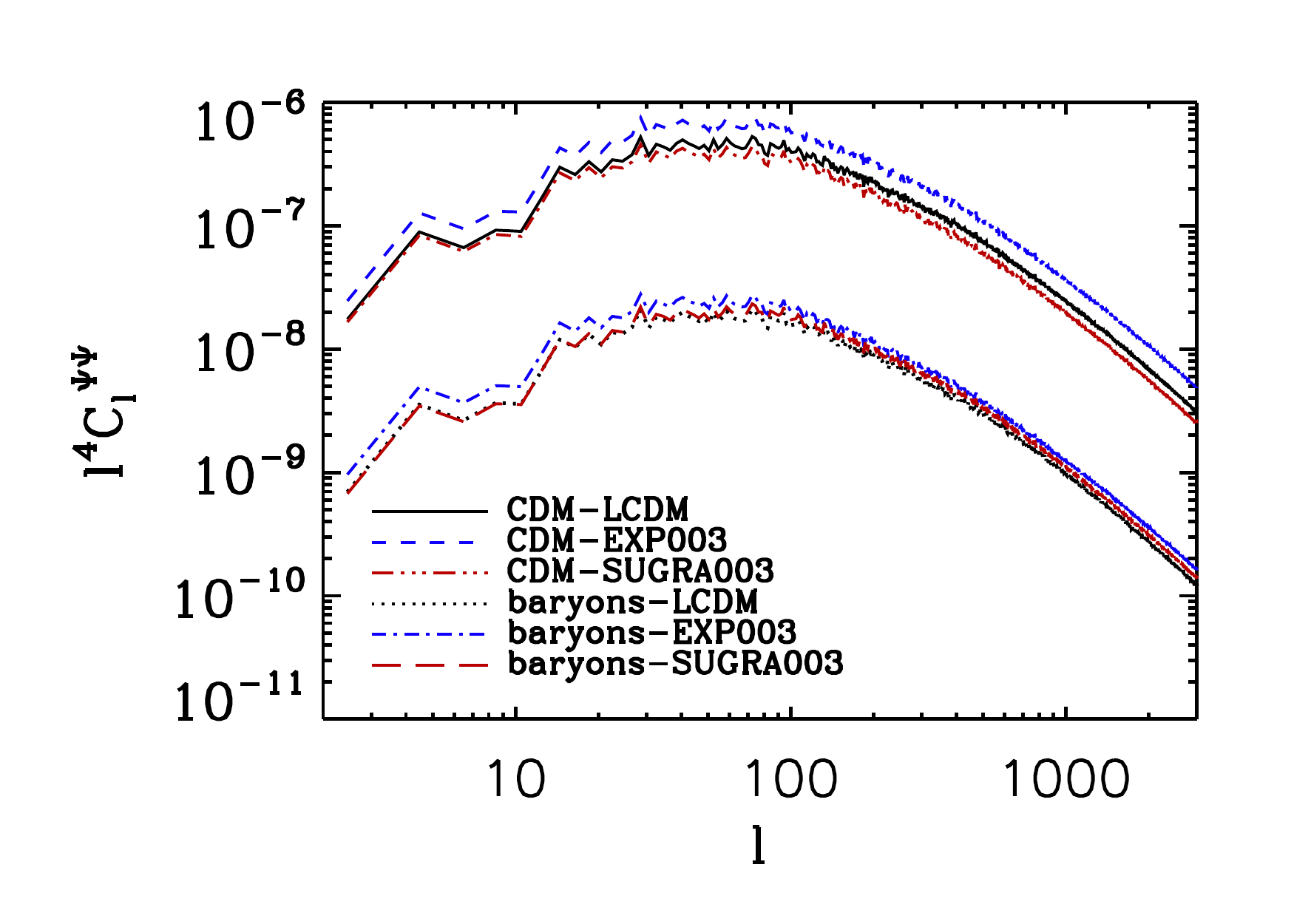}
\caption{Lensing potential power spectra of the baryon and CDM
  components, for the models considered in the text.} 
\label{fig:baryon_potentials}
\end{center}
\end{figure*} 
For these models, the pattern across the sky is similar in morphology, 
reflecting the same initial conditions in the {\small CoDECS} simulations. The
amplitude of the lensing signal, though, depends on the
integral over the subsequent structure formation epoch, weighted with the
lensing cross section, $(r_{*}-r)/(r_{*}r)$, in
Eq.~(\ref{lensingpotential}).  Already at the
level of the maps, the increase in the CMB weak lensing power for the
EXP003 model can be seen when looking at the maximum value assumed by $|{\bf
  \alpha}|$ in the lower left panel of Fig.~\ref{fig:exp003maps}, as
compared to the same quantity in Fig.~\ref{fig:lcdmmaps}, and
reflects the behaviour of structure formation, 
affected by the different growth factor and density perturbation
amplitude with respect to the $\Lambda$CDM
case. On the other hand, the amplitude of the lensing effects for the case
of SUGRA003 becomes closer again to the $\Lambda$CDM model, although as
we discuss below, a
finite and macroscopic difference remains also in this case,
reflecting the markedly different background and dynamic evolution
at the relevant epochs for CMB lensing.

More quantitatively, the mean value of the deflection angle modulus in
the simulated maps is given by ${\rm mean}(|{\bf \alpha}|)= 1.75',
2.12', 1.66'$, and the standard deviation is equal to 
$\sigma(|{\bf \alpha}|)=0.92',1.11', 0.87'$, for the adopted models
$\Lambda$CDM, EXP003 and SUGRA003, respectively.
For comparison, a synthetic
Gaussian map, produced with the lensing potential power spectrum
generated with CAMB (i.e. taking into account also the scales not
covered by the simulation box size)
using a WMAP-7 $\Lambda$CDM cosmology  
\cite{komatsu_etal_2011}, is characterised by 
$\sigma(|{\bf \alpha}|)=1.2'$; therefore the lack of scales larger
than the simulation box results in a suppression in $\sigma(|{\bf \alpha}|)$ of
about $23\%$ in the simulated CMB lensing signal. 

As explained in \S \ref{sec:codecs}, in the {\small CoDECS} simulations, baryon particles
are assumed to be collisionless. Nonetheless, since baryons are uncoupled from dark energy 
for the cDE models considered in this work, we expect that the relative behaviour of these models, 
with respect to the $\Lambda$CDM case, would be consistent with the baryon collisionless case, 
even when the effects of baryon physics on the matter power spectrum \cite{Semboloni_etal_2012} were correctly taken into account.

Therefore, in Figs.~\ref{fig:lcdmmapsbcdm}-\ref{fig:sugra003mapsbcdm}
we show collisionless baryon contributions to the 
quantities relevant for CMB lensing. We observe that baryons mostly follow the dynamics of
CDM particles, which are the dominant non-relativistic component.
For this reason, the morphology of the lensing potential and
deflection angle maps is markedly similar between baryons and CDM, and the general trend,
described above for the total matter contribution, is reproduced
individually by each of the components. On the other hand, the
contributions to the signal amplitude depend on the
relative abundances. In particular, for $\Lambda$CDM, EXP003 and
SUGRA003, respectively, baryons contribute to 
${\bf\alpha}$ by ${\rm mean}(|{\bf \alpha}|)= 0.29', 0.34', 0.30'$, and
$\sigma(|{\bf \alpha}|)=0.15',0.18', 0.16'$, while CDM particles give
${\rm mean}(|{\bf \alpha}|)= 1.46', 1.78', 1.36'$, and 
$\sigma(|{\bf \alpha}|)=0.76',0.93', 0.71'$. Therefore, baryons
produce only $\sim 1/3$ of the total rms of the
deflection angle modulus. 
\begin{figure*}
\begin{center}
\includegraphics[width=0.8\textwidth]{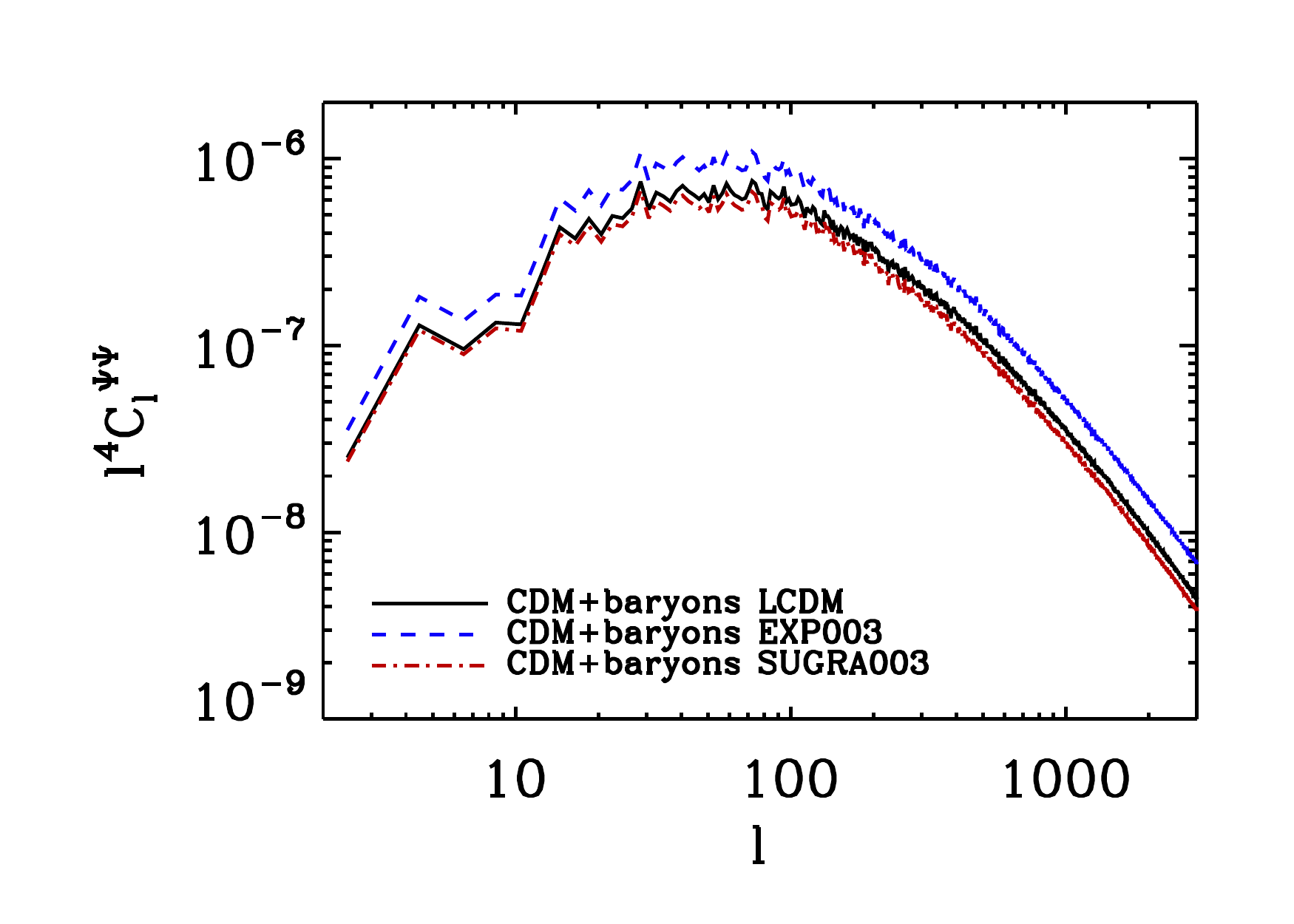}
\includegraphics[width=0.8\textwidth]{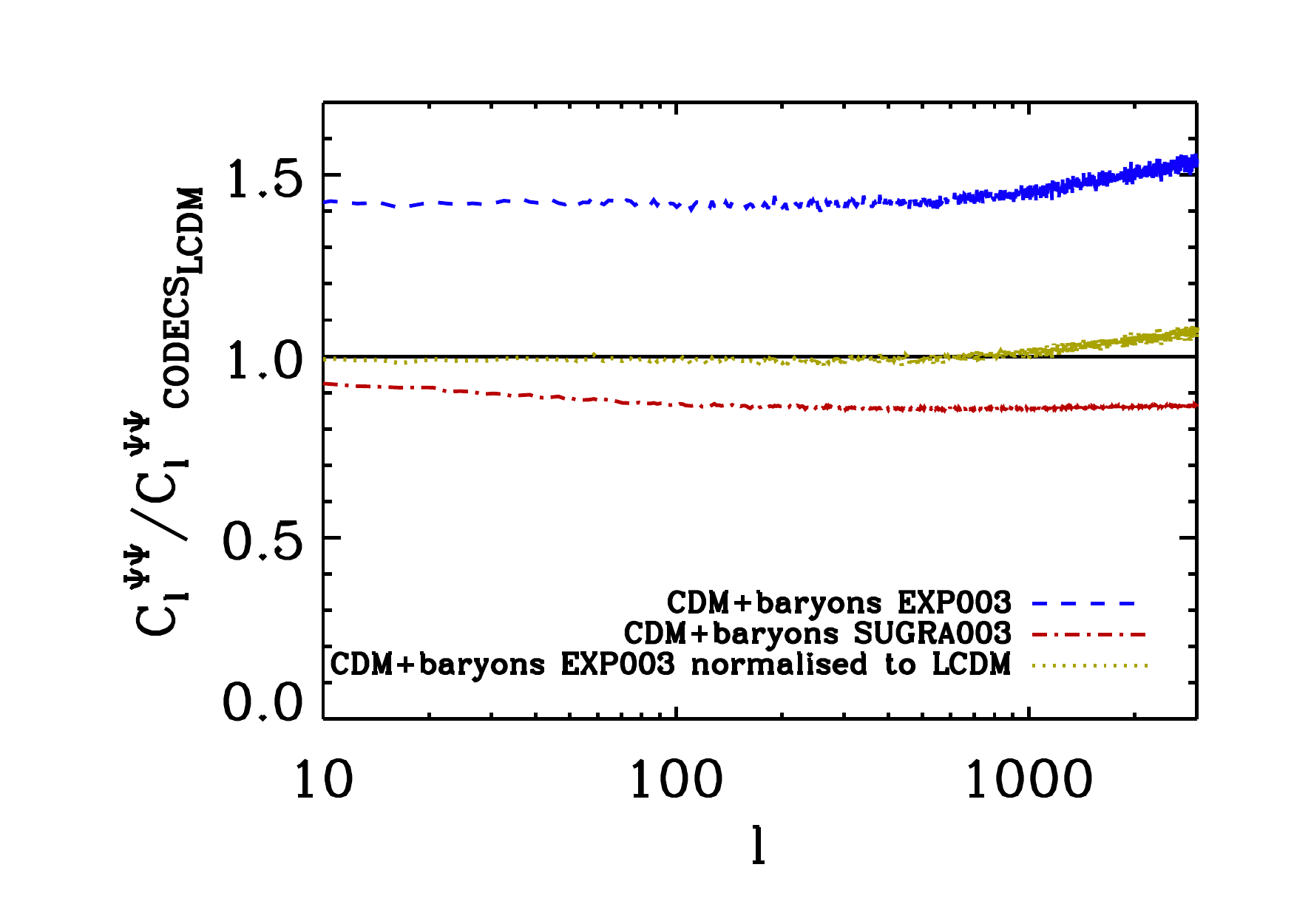}
\caption{Upper panel: lensing potential angular power spectra for the models considered in
  the present work. Lower panel: ratios of the simulated EXP003 and SUGRA003 lensing potential
  power spectra with respect to the $\Lambda$CDM one. The EXP003
  signal is also shown after
  re-scaling to the matter power normalisation of the $\Lambda$CDM model at present,
as explained in the text.}
\label{fig:lensing_potentials}
\end{center}
\end{figure*}
\begin{figure*}
\begin{center}
\includegraphics[width=0.8\textwidth]{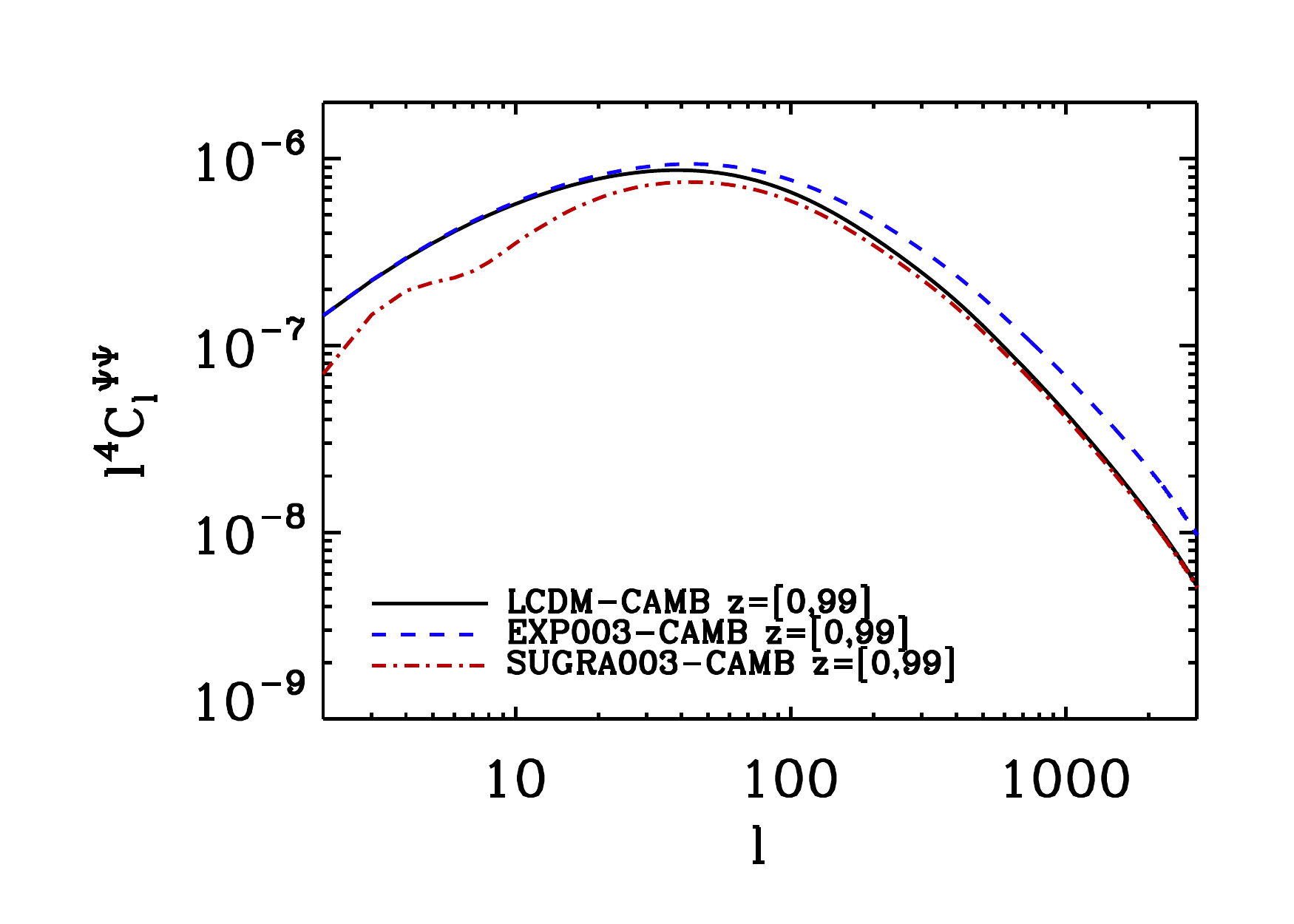}
\caption{Semi-analytic expectations of the lensing
  potential angular power spectra for the $\Lambda$CDM,
  EXP003, and SUGRA003 models, obtained with a modified CAMB version,
  as described in 
  the text. }
\label{fig:lensing_potentials_vale}
\end{center}
\end{figure*}
This can be seen, for example, by looking at
Fig.~\ref{fig:baryon_potentials}: here we plot the lensing potential
produced by both baryons and CDM components, for the three models discussed
above in the text. First, we see that there is a difference in power
of order $\sim \mathcal O(1)$ between the CDM and the baryon spectra. 
Moreover, the trends of the EXP003 lensing potential $C_l$, which are
higher than the corresponding spectra in the $\Lambda$CDM model at all
scales, both for 
the CDM and baryon cases (dashed and 
dot-dashed lines, respectively, in Fig.~\ref{fig:baryon_potentials}),
are well expected, 
given the higher matter power spectrum normalisation at the present epoch
with respect to the $\Lambda$CDM cosmology.\\ 
On the contrary, for the SUGRA003 model, baryons and CDM have a different and
unexpected trend with respect to the $\Lambda$CDM and EXP003
ones. Indeed, for SUGRA 
cosmologies, the CDM lensing potential power (3-dot--dashed line) lies below
the $\Lambda$CDM one (solid line),
while the baryon lensing potential power (long-dashed line) is slightly
above the $\Lambda$CDM one (dotted line).\\
In all the three cases, the differences among the 
baryon lensing potential power spectra are
visibly less marked than for the CDM components. Therefore, even if
collisionless, baryon dynamics does not perfectly follow the CDM
dynamics. The different behaviours between the baryon and CDM
  cases, in the three cosmologies considered, are due to the DE
  coupling with CDM, which is absent for baryons. 
In addition, at the end of this Section, we give a physical
explanation to the unexpected 
CDM lensing potential trend for the SUGRA003 case.

In the upper panel of Fig.~\ref{fig:lensing_potentials} we show the
total lensing 
potential power spectra extracted from the simulations, for the
$\Lambda$CDM, EXP003 and SUGRA003 models. 
Several features can be noticed in this figure; we start with
comments related to the properties of the simulations, regardless  
of the cosmological model, and then we describe the
different features reflecting the underlying cDE models.\\
First, the scale in which {\small CoDECS} allow to trace real
structures in the 
lensing pattern is broader by a factor of $\sim 2$ with respect to the
results of \citep{carbone_etal_2008,carbone_etal_2009}, obtained via
ray-tracing across the Millennium Run simulation. In particular, we
have checked that, in the $\Lambda$CDM case, {\small CoDECS} reproduce
correctly the lensing potential power on scales $l>200$. This is
due to its box side, which is doubled with respect
to the Millennium Run, able to reproduce the lensing
signal only for $l>400$.\\
Second, the plot shows how the lensing power faithfully traces
differences in structure formation at the epoch in which the lensing
cross section for CMB lensing is maximum, i.e. $z\simeq 1\pm 0.5$, as
anticipated in earlier works \citep{acquaviva_baccigalupi_2006}. This
effect is described by the different amplitudes and shapes of the
lensing potential power spectra as represented by the solid black line,
the dashed blue line and the dot-dashed red line, for the
$\Lambda$CDM, EXP003 and SUGRA003 models, respectively.

The upper panel of Fig.~\ref{fig:lensing_potentials} has to be
compared with Fig.~\ref{fig:lensing_potentials_vale},
in which we show the semi-analytical predictions obtained with the
implementation of 
the publicly available Code for Anisotropies
in the Microwave Background (CAMB\footnote{See http://camb.info}),
adapted to interacting dark energy cosmologies \cite{Pettorino_etal_2012}.\\ 
From this comparison, we notice that,
on the scales probed by the simulations, $l>200$, there is 
a qualitative good agreement between semi-analytical results and
simulated data, but, except for the $\Lambda$CDM case, already at multipoles
$l>400$, where the non-linear 
structure evolution starts to affect the lensing signal, the
semi-analytical curves deviate from the simulated ones, due
to the Halofit \citep{smith_etal_2003} non-linear implementation
present in CAMB, which has been tested against N-body
simulations only in $\Lambda$CDM cosmologies.\\
We avoid to
compare in the same plot the semi-analytical
expectations with the simulated results from {\small CoDECS}. 
In fact, the integration of background equations for coupled quintessence models requires an
iterative routine that allows to tune the initial conditions such that
the wished background cosmological 
parameters at present are recovered. For the SUGRA models, the presence of the bounce increases
the level of fine tuning and makes the background evolution more unstable with respect to small
variations of the coupling and potential parameter. As a consequence, the background values
obtained in the SUGRA003 model for CAMB are not exactly the same as
the ones used for the {\small CoDECS} background. For this reason, in
order to compare the semi-analytical expectations of the
lensing potential among the three cosmological models considered in
this work, and check if they confirm qualitatively the $C_l$ trends extracted
directly from {\small CoDECS}, we have decided to produce the
corresponding $\Lambda$CDM 
and EXP003 spectra using exactly the same background values 
chosen in CAMB for SUGRA003, instead of adopting the {\small CODECS}
background values.

In the lower panel of Fig.~\ref{fig:lensing_potentials}, we show the
ratios of the lensing potential spectra with respect to the
$\Lambda$CDM  signal, as extracted from the 
simulations, and, for the EXP003 case,  
we also display the same ratio normalised to $\sigma_{8(\Lambda{\rm
    CDM})}$, obtained dividing the EXP003 lensing potential $C_l$ by
$\sigma_{8({\rm EXP003})}^2$ and multiplying it by $\sigma_{8(\Lambda{\rm
    CDM})}^2$ (see
Table~\ref{tab:models}). In particular, this model shows a power
excess on all scales $l>200$ of $\sim 50 \%$ with respect to the
$\Lambda$CDM spectrum (dashed blue line in
the lower panel of Fig.~\ref{fig:lensing_potentials}). As we can
observe from the dotted green line in the same plot, this excess of
power is mostly due to the different matter power spectrum
normalisation at $z=0$ between the $\Lambda$CDM and EXP003 cosmologies
arising as a consequence of the enhanced linear growth.
Nonetheless, on non-linear scales, $l>1000$, a residual excess is present,
produced by the different 
non-linear dynamics and evolution of the CDM perturbations, given the matter
coupling with the DE. At $l\sim 3000$ non-linearities contribute to the
power excess in the EXP003 lensing signal by $\lessapprox 10\%$. This
effect can not be reproduced with the use of semi-analytical
techniques in cDE models,
being the product of the LSS non-linear evolution accurately
described only via N-body simulations.
Furthermore, it would be misleading to conclude that the $\sigma_8$
degeneracy between $\Lambda$CDM and EXP003 breaks only beyond the arcminute
scale, i.e. where the non-linear structure evolution is more
efficient, especially due to the coupling between DE and CDM
\citep{Baldi_2012a}. This is what one could wrongly infer owing to the
limited box size of the adopted N-body simulations. In fact, having a
look at the 
semi-analytic predictions in Fig.~\ref{fig:lensing_potentials_vale},
which reproduce the lensing signal on scales $l<200$, it
is easy to observe 
that the $\sigma_8$ degeneracy between the EXP003 and $\Lambda$CDM
cosmologies is completely broken if 
also scales larger than the simulation box
are taken into account. This feature will be transferred,
after lensing, to the first CMB temperature and polarisation
peaks, allowing to clearly distinguish between the two models even
after normalisation to the same $\sigma_8$. Lensed T, E, and B
CMB spectra in cDE cosmologies will be accurately treated in a future work.

Finally, as the dot-dashed red lines in the upper and lower panels
of Fig.~\ref{fig:lensing_potentials} clearly show, and as anticipated
above, the lensing effect produced by the SUGRA003 model shows a
completely opposite trend with respect to the EXP003 one. In fact, in this
case, the combination of modified geometry and abundance of lenses  
determines a loss of power, remarkably at the
$10\%$ level on all the scales spanned by the simulations. We have also checked that, 
as a consequence of SUGRA003 being similar to $\Lambda$CDM in  
perturbation power at present, this effect 
stays unchanged after $\sigma_8$ normalisation. 
\begin{figure*}
\begin{center}
\resizebox{7cm}{!}{\includegraphics{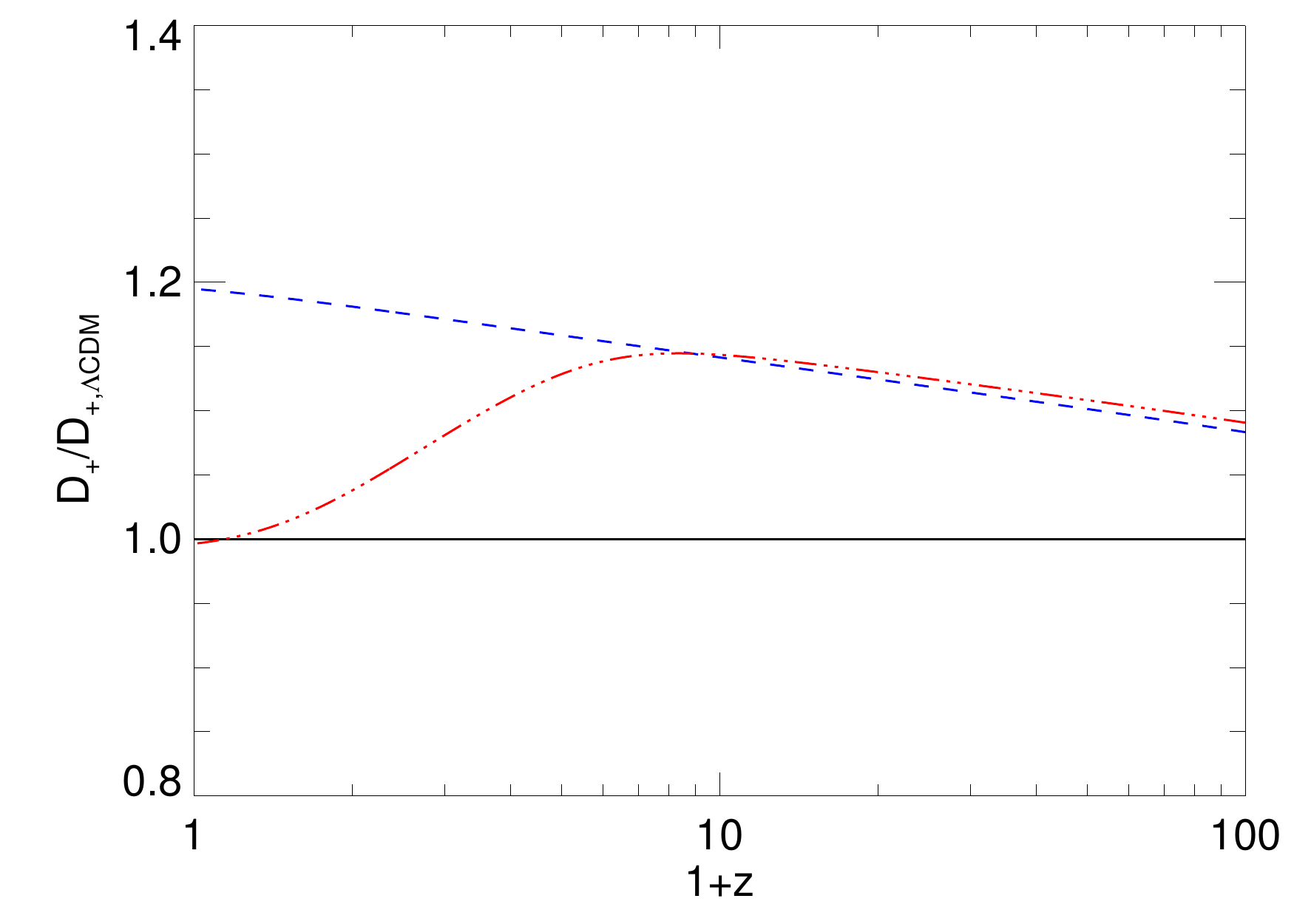}}
\resizebox{7cm}{!}{\includegraphics{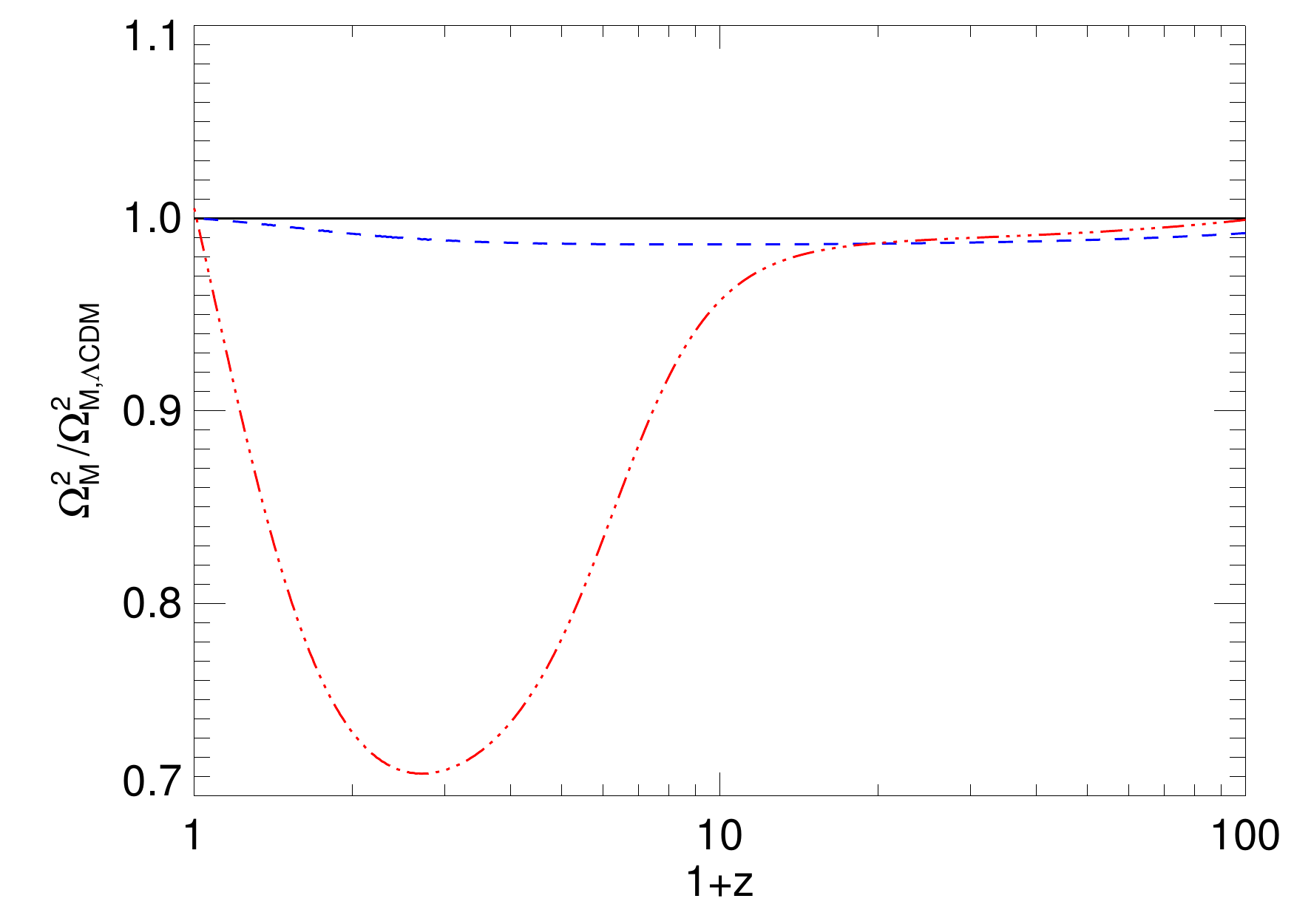}}
\resizebox{7cm}{!}{\includegraphics{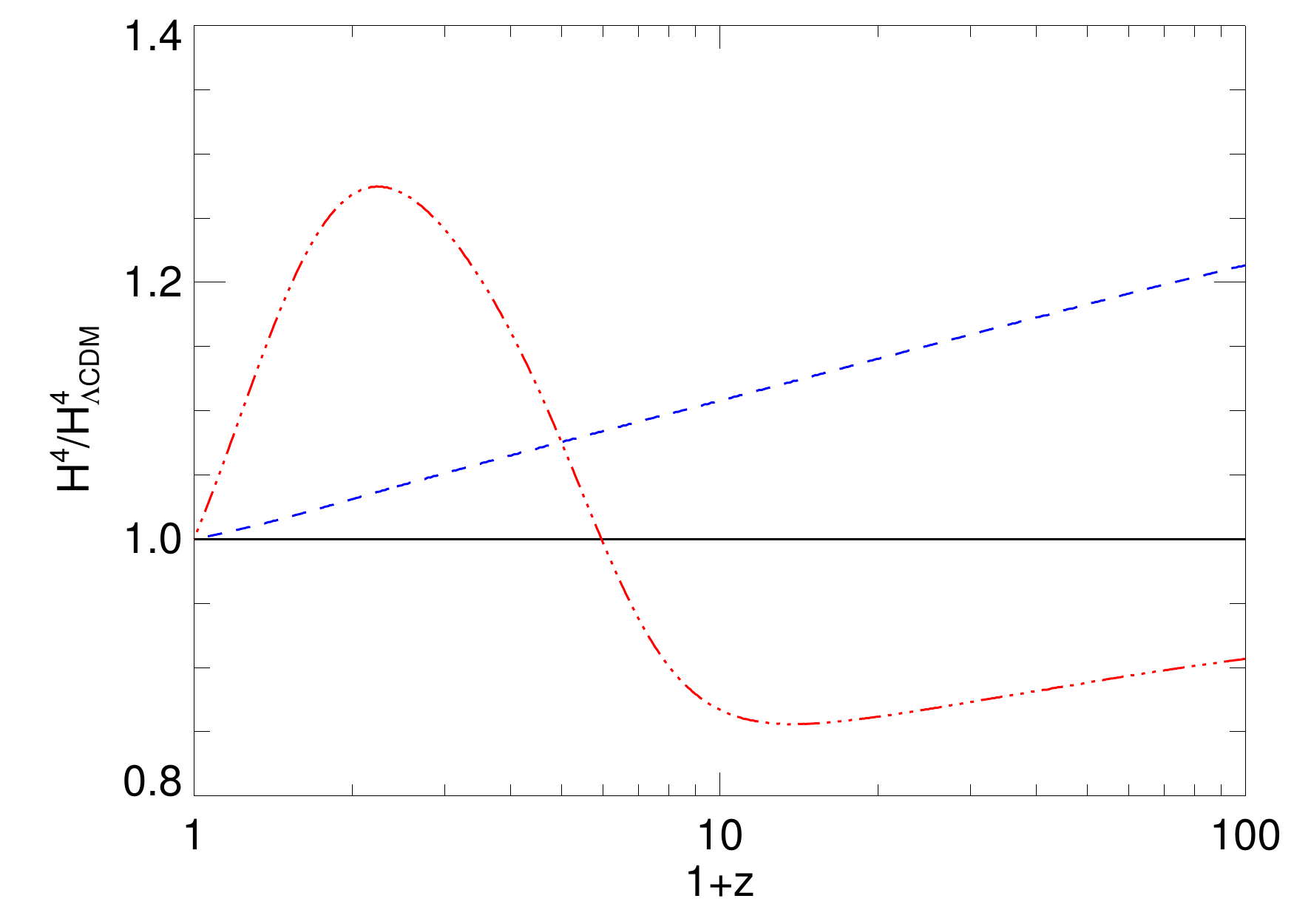}}
\resizebox{7cm}{!}{\includegraphics{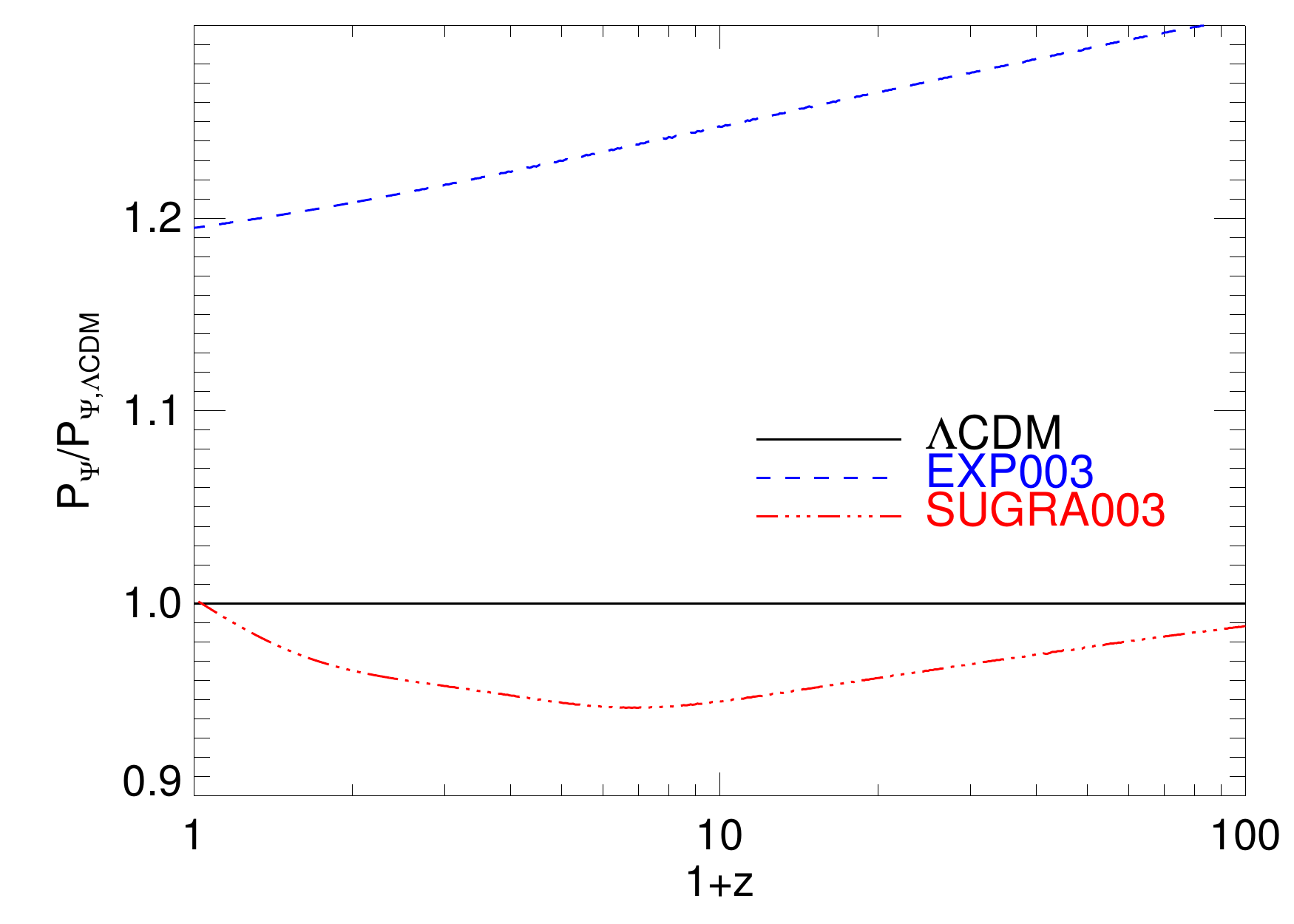}}
\caption{The redshift evolution of the three terms entering the expression for the lensing power spectrum amplitude, and their combination, as compared to the $\Lambda $CDM case. As one can see from the plots, the different terms have very different evolutions that partly compensate with each other, resulting in an overall evolution of the lensing power amplitude as compared to $\Lambda $CDM displayed in the lower-right panel. The latter clearly shows how an enhanced lensing signal is expected for the EXP003 model, while a moderately suppressed lensing power is predicted for SUGRA003, despite for this model the amplitude of matter density perturbations is always larger than in $\Lambda $CDM.}
\label{fig:interpretation}
\end{center}
\end{figure*}
Such phenomenology might appear counter-intuitive as the SUGRA003 model is expected to always have a higher amplitude of matter density perturbations as compared to $\Lambda $CDM, except for $z=z_{\rm CMB}$ and $z=0$ \citep[][]{CoDECS}. One would intuitively expect also a larger lensing potential arising as the integral over the line-of-sight of a density power spectrum with a higher amplitude than the standard $\Lambda $CDM cosmology. However, this is not the case due to the interplay between the evolution of background quantities and density perturbations in determining the lensing power spectrum $\mathcal{P}_{\Psi }$. In fact, within the small-scale Limber approximation, the latter is related to the matter density power spectrum through the relation:
\begin{equation}
\mathcal{P}_{\Psi }(k\,; a) = \frac{9\Omega _{\rm M}^{2}(a)H^{4}(a)}{8\pi ^{2}}D_{+}(a)\frac{P(k)}{k}\,,
\end{equation}
such that the excess with respect to the standard $\Lambda $CDM cosmology can be written as the combination of three main factors:
\begin{equation}
\label{limber_ratio}
\frac{\mathcal{P}_{\Psi }}{\mathcal{P}_{\Psi\,, \Lambda }} 
=\left[ \frac{\Omega _{{\rm M}}(a)}{\Omega _{{\rm M}\,,\Lambda }(a)}\right] ^{2} \left[\frac{H(a)}{H_{\Lambda }(a)}\right]^{4} \left[\frac{D_{+}(a)}{D_{+\,,\Lambda }(a)}\right] \,,
\end{equation}
where $D_{+}(a)$ is the linear growth factor normalised at recombination, and where we have taken into account that all the cosmological models under investigation share the same linear matter power spectrum at last scattering. 

These three different terms are shown in Fig.~\ref{fig:interpretation}, together with their combined effect on the lensing power spectrum amplitude. While the evolution of the growth factor alone would suggest a higher lensing efficiency due to the larger amplitude of
density perturbations, the remaining two terms compensate for this effect in a non-trivial way, determining an overall evolution of the expected lensing power amplitude fully consistent with the results obtained through our ray-tracing procedure.
Clearly, such estimate is expected to fail at highly nonlinear scales,
where in fact our fully nonlinear treatment is capable to capture
additional features that are not accurately predicted by the ratio of
Eq.~(\ref{limber_ratio}). 

\section{Conclusions}
\label{sec:conclusions}
The modification of the anisotropies in the CMB induced by forming  
cosmological structures through gravitational lensing is one of the
most important topics in modern  
cosmology. The lensing signal probes the expansion rate, and thus the
abundances of the dark cosmological  
components, including DE, through the distribution
of lenses at an epoch which corresponds  
to the onset of cosmic acceleration. 

In view of the existing plans for ground and space  based wide field and deep
observations of LSS, large N-body simulations are being produced or
planned, involving DE models as well as scenarios in which  
gravity is modified, in order to simulate with high accuracy the
cosmic acceleration in theories alternative to the
cosmological constant scenario. CMB anisotropy data, on the other hand, are
available over the entire sky, with angular resolution
reaching a few arcminutes, and 
sensitivity of a few micro-Kelvin, in total  
intensity and polarisation. 

In this context, it is important to produce simulations of
the expected CMB lensing pattern in the aforementioned scenarios, for
two reasons. 
A first one concerns testing and
developing our capability of implementing  
N-body based simulations of CMB lensing, involving ray tracing through
simulated structures. A second one  
consists in studying and quantifying the sensitivity of CMB lensing
variables to the underlying cosmological  
model, concerning the modelling of DE, in particular.
 
In this work, we progressed on both aspects, by simulating the CMB weak lensing potential
through N-body simulations featuring various DE models characterised
by modified expansion histories, modified cosmological perturbation dynamics, and  
the presence of a coupling between CDM particles
and DE, in comparison to the expectations of the $\Lambda$CDM
model. We constructed suitable grids of the 
gravitational potential output from N-body simulations,  
and implement ray tracing in the Born approximation. We derived maps of
the deflection angle and lensing potential, verifying our
capability of tracing N-body structures in the  
CMB lensing signal on a range of scales going from the arcminute to
the degree scale, dictated, as expected, entirely by the simulation  
resolution and extension. 
Following well known geometrical properties of CMB lensing, which has its cross section
maximum at redshifts of about $z\simeq 1\pm 0.5$, we studied how  
the simulated lensing signal traces differences in the background and
perturbation dynamics at the corresponding epoch.  
We considered two DE models, EXP003 and SUGRA003, the first featuring an increased structure
formation rate with respect to the $\Lambda$CDM  
at all epochs, and the second characterised by a matter perturbation normalisation close to the
$\Lambda$CDM at the initial and present epochs,  
being different only at intermediate redshifts. We show how the lensing variables faithfully record the
differences of the models with respect to the $\Lambda$CDM scenario  
at the onset of acceleration, in particular for the second scenario,
degenerate with the $\Lambda$CDM at present and at the  
initial time. Correspondingly to the different amplitudes and dynamics
of the distribution of lenses in the various models, for the SUGRA003
cosmology the lensing
potential power results to be of order $10\%$ smaller than the
corresponding $\Lambda$CDM signal on all
the scales covered by the simulation. This is a quite unexpected
result, and it is due to the
compensation between the larger power of the matter perturbations and
the suppressing effect of the background and matter density evolution characterising this
specific class of cDE models.
The {\small CoDECS} simulations store the gravitational potential of baryons and CDM  
particles separately;
therefore, we inspected the results for the baryon component only,  
finding a substantial compensation as for the CDM particles, with the  
power being
slightly higher than $\Lambda$CDM in this case. We attributed this  
occurrence to the coupling with DE, which concerns CDM particles, and  
not baryons. 
For the EXP003 model, we find an opposite result, i.e. we find an
excess of order $10\%$, with respect to the $\Lambda$CDM case, for the
lensing potential power on multipoles 
$l>1000$, after normalisation of the signals to the same
$\sigma_8$. On the large scales, i.e. on multipoles $l<20$ instead,
the semi-analytical expectations predict that the
signals tend to converge to the same amplitude independently of the
underlying cosmology, owing to the same
curvature perturbation power amplitude assumed for the initial
conditions of the N-body simulations. This means that the $\sigma_8$
degeneracy between the $\Lambda$CDM and EXP003 model can be broken if a
large interval of scales is analysed in the observations.

These results confirm the relevance of CMB lensing as a probe for DE
at the early stages of cosmic acceleration, and demonstrate the  
reliability of CMB lensing based on the existing N-body simulations,
in the context of DE studies. At the same time, our study  
clearly indicates paths forward, especially in terms of box size and
alternative simulated cosmologies, in order to be able to access super-degree  
angular scales, as well as accurately mapping the lensing pattern on
arcminute and sub-arcminute angular scales; the latter in particular  
represents a crucial regime for CMB lensing, as it gets to dominate
with respect to the decaying CMB anisotropies of primordial origin  
due to diffuse damping at last scattering, and at the same time
involving non-linear physics where semi-analytic modelling may be
either unavailable or inaccurate. In this respect, N-body simulations of actual
structure formation may be the only way to well characterise DE
effects in such a regime.

\section{Acknowledgements}
We acknowledge the use of the publicly available Code for Anisotropies
in the Microwave Background\footnote{CAMB, see camb.info.},  
and the Hierarchical Equalised Latitude Pixel spherical pixelisation
scheme (HEALPix\footnote{http://healpix.jpl.nasa.gov}).
The ray-tracing computations have been performed on the IBM PLX-GPU cluster at
CINECA (Consorzio Interuniversitario del Nord-Est per il
Calcolo Automatico), Bologna, with CPU time assigned under a CINECA
class-C call. Part of the
computation of this paper was done on the Andromeda Cluster at the
University of Geneva.
CC acknowledges the INAF Fellowships Programme 2010.
MB is supported by the Marie Curie Intra European Fellowship ``SIDUN"
within the 7th Framework Programme of theEuropean Community. MB also
acknowledges partial support  by the DFG Cluster of Excellence ``Origin and
Structure of the Universe'' and by the TRR33 Transregio Collaborative
Research Network on the ``Dark Universe''. VP is supported by Marie Curie IEF, Project DEMO. CB acknowledges support by the INFN PD51 initiative and the Italian Space
Agency through the ASI contracts Euclid-IC (I/031/10/0).

\end{document}